\documentclass{article}
\usepackage{latexsym}
\usepackage{color,graphicx}
\usepackage{latexsym,amssymb,epsf} 

\newcommand{\bge}{\begin{equation}}
\newcommand{\ee}{\end{equation}}
\newcommand{\bgc}{\begin{center}}
\newcommand{\ec}{\end{center}}
\newcommand{\bgea}{\begin{eqnarray}}
\newcommand{\eea}{\end{eqnarray}}
\newcommand{\bgeas}{\begin{eqnarray*}}
\newcommand{\eeas}{\end{eqnarray*}}

\begin{document}

\title{
  \bf  Two-dimensional hydrodynamic lattice-gas simulations of binary immiscible and ternary amphiphilic fluid flow through porous media}

\author{
  J.-B. Maillet$^1$\\
  {\footnotesize Schlumberger Cambridge Research, High Cross,
    Madingley Road,}\\
  {\footnotesize Cambridge, CB3 0EL, U.K.}\\
  Peter V. Coveney,\\
  {\footnotesize Centre for Computational Science,
    Queen Mary and Westfield College,}\\
  {\footnotesize University of London, Mile End Road,}\\
  {\footnotesize London E1 4NS, U.K.}\\
  {\footnotesize{\tt p.v.coveney@qmw.ac.uk}}\\[1.5cm]
  {\footnotesize $^1$ Present address: CECAM, Ecole Normale
    Sup\'{e}rieure,}\\
  {\footnotesize 46, All\'ee d'Italie, 69364 Lyon Cedex 07,
  France.}\\[0.3cm] 
}

\maketitle

\newpage

\begin{abstract}
The behaviour of two dimensional binary and ternary amphiphilic fluids
under flow conditions is
investigated using a hydrodynamic lattice gas model. After the validation of the
model in simple cases (Poiseuille flow, Darcy's law for single
component fluids),
attention is focussed on the properties of binary immiscible fluids in
porous media. An extension of Darcy's
law which explicitly admits a viscous coupling between the fluids is
verified, and evidence of capillary effects are described.
The influence of a third component, namely surfactant, is studied in the
same context.\\
Invasion simulations have also been performed. The effect of the
applied force on the invasion process is reported. As the forcing
level increases, the invasion process becomes faster and the residual
oil saturation decreases.
The introduction of surfactant in the invading phase during imbibition
produces new phenomena, including emulsification and micellisation. At very low
fluid forcing levels,
this leads to the production of a low-resistance gel, which then slows down the progress
of the invading fluid. At long times (beyond the water percolation threshold),
the concentration of remaining oil within the porous medium is lowered by the
action of surfactant, thus enhancing oil recovery.
On the other hand, the introduction of surfactant in the invading
phase during drainage simulations slows down the invasion process  - the
invading fluid takes a more tortuous path to invade the porous
medium - and reduces the oil recovery (the residual oil saturation
increases).
\end{abstract}
\newpage

\vskip 20pt

\section{Introduction}
Since the ability of lattice gas automaton (LGA) models to reproduce correctly the
incompressible Navier-Stokes equations was established~\cite{frisch,wolfram},
these models have been intensively studied. Rothman and Keller
developed an extension of the one-component model for simulating
binary immiscible
fluids~\cite{rothman}. The introduction of a third component, namely surfactant, is due to
Boghosian \emph{et al.}~\cite{boghosian}. The surfactant particle acts as a
point dipole, and tends to stay at the interface between the two immiscible fluids. It
can also form micelles, when its concentration exceeds a
particular value (the critical
micelle concentration).
This work logically follows the two dimensional studies performed  by
Wilson and Coveney on flowing multiphase fluids, including their
application to porous media~\cite{wilsoncoveney}. Some preliminary
results have been described in a first publication~\cite{coveneymaillet}. Complex fluid flow in porous media is both a scientifically challenging problem and a field of great practical importance, from oil and gas production to environmental issues in ground state water flows~\cite{bookrothman}.\\
The paper is structured as follow: a short description of the hydrodynamic lattice gas model is given in Section 2. We
present in Section 3 the results obtained for the simulation of single phase flow
through a 2D channel and the verification of some theoretical predictions.
These computations allow one to calculate the viscosity of the fluid.
Section 4 is devoted to the
verification of Darcy's law and to its generalisation to the case of
multiphase fluids. Invasion phenomena are investigated in Sections 5
and 6 and conclusions are presented in
Section 7.

\section{Description of the model}
\paragraph{}
According to the lattice gas model for
microemulsions~\cite{boghosian}, the sitewise interaction energy of
the system can be written:
\begin{equation}
\Delta H_{int}=\alpha \Delta H_{cc} + \mu \Delta H_{cd} + \epsilon
\Delta H_{dc} + \zeta \Delta H_{dd}.
\end{equation}
These terms correspond respectively to the relative immiscibility of oil and water,
the tendency of surrounding dipoles to bend round oil or water
particles and clusters, the propensity of surfactant molecules to
align across oil-water interfaces and a contribution from pairwise
interactions between surfactant.\\
The use of a probabilistic, or Monte Carlo process, to choose the outgoing state
when particles collide leads to the introduction of a
temperature-like parameter $\beta$, which is, however, not related to a true
thermodynamic temperature. This parameter does not allow the analytical prediction of the
viscosity. In essentially all such lattice-gas models involving
interactions between multicomponent species, the
condition of detailed balance is not satisfied. This
leads to the fact that one cannot be sure, \emph{a priori}, that an equilibrium
state exists. Nevertheless, numerical simulations confirm
that steady states are reached. 
In this paper, the following set of parameters have been used:
\begin{center}
\begin{tabular}{ccc}
$\alpha$ &=& $1.0$\\
$\mu$ &=& $0.001$\\
$\epsilon$ &=& $8.0$\\
$\zeta$ &=& $0.005$\\
$\beta$ &=& $1.0$
\end{tabular}
\end{center}
\vspace{3mm}
We make use of a triangular (FHP) lattice, with six directions at each site.
There can be up to seven particles at each site. The reduced density of
a fluid phase is defined as the average number of particles of this
fluid phase (colour)
per lattice site, divided by $7$ (six lattice directions and one rest particle).
Note that all simulations performed in this paper are two dimensional.
A three dimensional version of the model has been
formulated~\cite{boghosian99} and similar investigations are already
underway using this high performance computing code~\cite{maillet}.
The different fluid forcing methods are described in a
previous paper~\cite{coveneymaillet}.\\
Lattice sites are selected at
random, on which momentum is added, so that either (a) the total
average momentum is kept constant (``pressure'' condition), or (b) the
total added momentum is constant (``gravity'' condition).
In order to study the process of fluid invasion into porous media,
some modifications have been made to our existing lattice-gas code~\cite{coveneymaillet}:
The simulation cell system is no longer periodic in the flow
direction (the vertical or y-direction), but retains periodicity in the x-transverse direction.
To achieve this, ``invisible'' rows at the top and the bottom of the lattice have
been added in order to simulate infinite columns of bulk oil and water
respectively.
As the total number of particles is conserved, they are wrapped
from top to bottom and \emph{vice versa}, but in so doing, they change their colour
to that of the bulk surrounding colour fluid.
When obstacle sites are present in the lattice, no-slip boundary conditions
are used, corresponding to a zero velocity condition at the boundary
in a conventional Navier-Stokes fluid. Obstacles may be
given a colour charge, thus assigning wettability properties to the
simulated rock species. The wettability index can vary over the range
$\{-7;+7\}$, $-7$ corresponding to a rock site \emph{full} of water (blue)
particles, $+7$ to a rock site \emph{full} of oil (red) particles
(i.e. maximally hydrophilic and hydrophobic respectively).

\section{Two-dimensional channel simulations}
\subsection{Single phase fluids}
In this section, we first check some basic properties of single phase
flow within channels in two dimensions, and then move on to consider
binary immiscible fluid flow.
\subsubsection{Velocity profile measurements}
\begin{figure}[ht]
\begin{center}
\scalebox{0.8}{\includegraphics{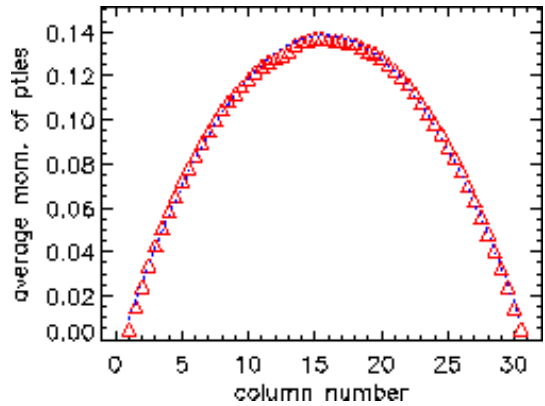}}
\end{center}
\caption{\textbf{Velocity profile averaged over 20000 timesteps for a reduced
density of 0.5 on a $32 \times 32$ lattice. The simulated points are
represented by triangles and the fitted curve is the dotted line. The
forcing level is 0.00075 using gravity conditions.}}
\label{velprofile32}
\end{figure}

We are concerned here with the flow of a single phase fluid through a
pipe (Poiseuille flow); the velocity profile in this case is known to
be parabolic. The results obtained using a $32 \times 32$ lattice are
displayed figure~\ref{velprofile32}. Also shown is a parabolic fit to
the curve. The agreement
between the simulated curve and the fit is good.\\
This lattice gas model is able to reproduce correctly the flow of a
single phase fluid through a pipe.
The velocity profiles for low-density fluids (0-2 particles per lattice
site) are less well-fitted by a parabola.
\subsubsection{Calculation of the viscosity}
\begin{figure}[ht]
\begin{center}
\scalebox{1.0}{\includegraphics{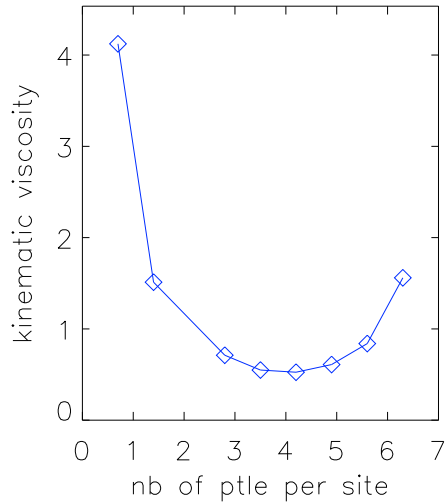}}
\end{center}
\caption{\textbf{Kinematic viscosity as a function of density for a $32 \times 32$ box size.}}
\label{viscosity32}
\end{figure}

From the velocity profile, we can extract the numerical value of the
maximum flow velocity which occurs in the centre of the pipe. The
relation between the velocity at the center of the pipe and the
kinematic viscosity is:
\begin{equation}
\nu=\frac{1}{8} \frac{FW}{Lg_{max}},
\end{equation}
where $\nu$ is the kinematic viscosity, $F$ is the forcing level, $W$
and $L$ are the width and the length of the channel respectively and
$g_{max}$ is the maximum of the velocity in the center of the pipe.
Carrying
out this calculation for different densities allow us to compute the viscosity of
the fluid as a function of density (figure~\ref{viscosity32}).\\
The result obtained with this model is in the same range as that
obtained by Kadanoff \emph{et al.}~\cite{kadanov}.

\subsection{Binary immiscible fluid flow in a channel}
\label{binaryflow}
Simulations of binary mixtures in a two-dimensional pipe were conducted on a
$32\times128$ lattice. The first set of simulations concerned the
influence of the wettability index on the shape of a
non-wetting fluid bubble and the second set involved a study of the
coupling between the two fluids.
\subsubsection{Influence of the wall wettability}
\begin{figure}[p]
\begin{center}
\scalebox{1.0}{\includegraphics{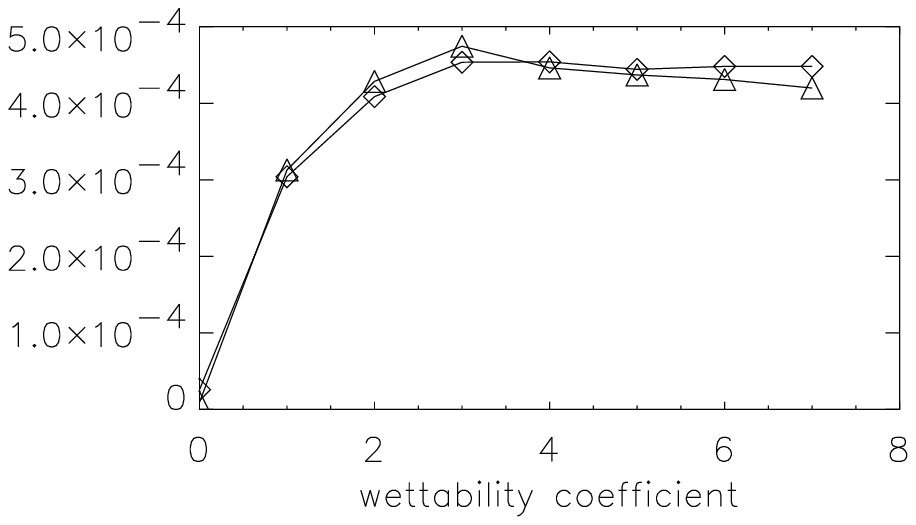}}
\scalebox{1.0}{\includegraphics{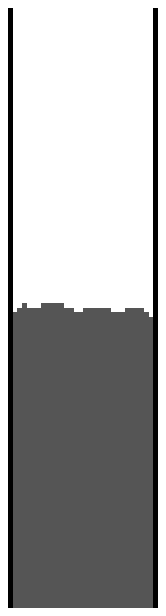}\includegraphics{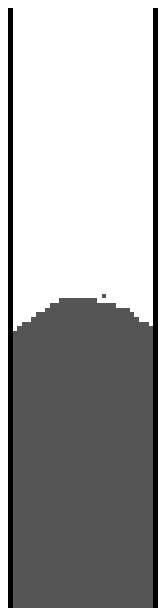}\includegraphics{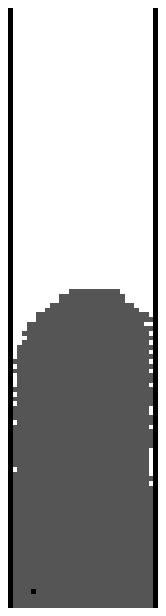}\includegraphics{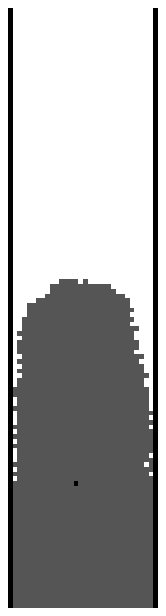}}
\end{center}
\caption{\textbf{Effect of the wettability index on the force needed
to keep the non-wetting fluid on the lattice. The curves with
diamonds and triangles
correspond to reduced densities of $0.35$ and $0.25$ respectively.
The configurations from left to right correspond to simulations with wettability $0$,
$1$, $2$, and $3$ (the wetting fluid is in white). The force is in
unit momentum per timestep and the lattice size is $32 \times 128$.}}
\label{capilar}
\end{figure}

In order to study the effect of the wettability index on the
shape of a non-wetting bubble, invasion conditions are used.
These conditions simulate infinite columns of wetting and non-wetting fluid at
the top and the bottom of the lattice respectively, thus allowing the non-wetting bubble to
adhere to the walls (at least at the bottom of the lattice). If the simulations are performed without bulk flow, the
wetting phase progressively invades the lattice (owing to capillary
effects), preventing a detailed study of the non-wetting bubble.
The simulations are run using pressure forcing applied to the non-wetting fluid, in order to achieve a state of zero flux (momentum is
added to the non-wetting phase until its flux reaches a zero value). The force
needed to keep the non-wetting
fluid on the lattice is then determined, as a function of the wettability of
the walls. Figure~\ref{capilar} displays the results for a 1:1 water
and oil mixture with a total reduced density of either 0.5 or 0.7. Each point is
averaged over 5 independent simulations, each of duration $20000$ timesteps.
Several points can be made. First of all, the curves corresponding to different reduced
densities are similar. They display a strong influence of the
wettability index in the range $0-3$, for which the force
needed to keep the non-wetting fluid on the lattice increases
strongly. For wettability indices greater than $3$, there is
no change in the restraining force. Snapshots at different times
during the simulation show that for non-wetting
walls, the interface is on average flat.
Increasing the wettability leads to deformation of
the interface, from a flat to a curved interface, and finally
to the detachment from the wall of the upper part of the non-wetting bubble, for an
index of $2$. Increasing the wettability index further has no effect
because the bubble has already detached from the wall.\\
It is surprising that the reduced densities of the fluids do not
influence this curve. One might have thought that the point at which the
bubble detaches from the wall would correspond to there being an equal number of particles per site
in the wetting fluid and in the obstacle sites so that, when the reduced
density of the wetting fluid increases, the bubble would detach
for greater wettability coefficients, but this is not found to be the case.

\subsubsection{Coupling}
\begin{figure}[ht]
\begin{center}
\scalebox{1.0}{\includegraphics{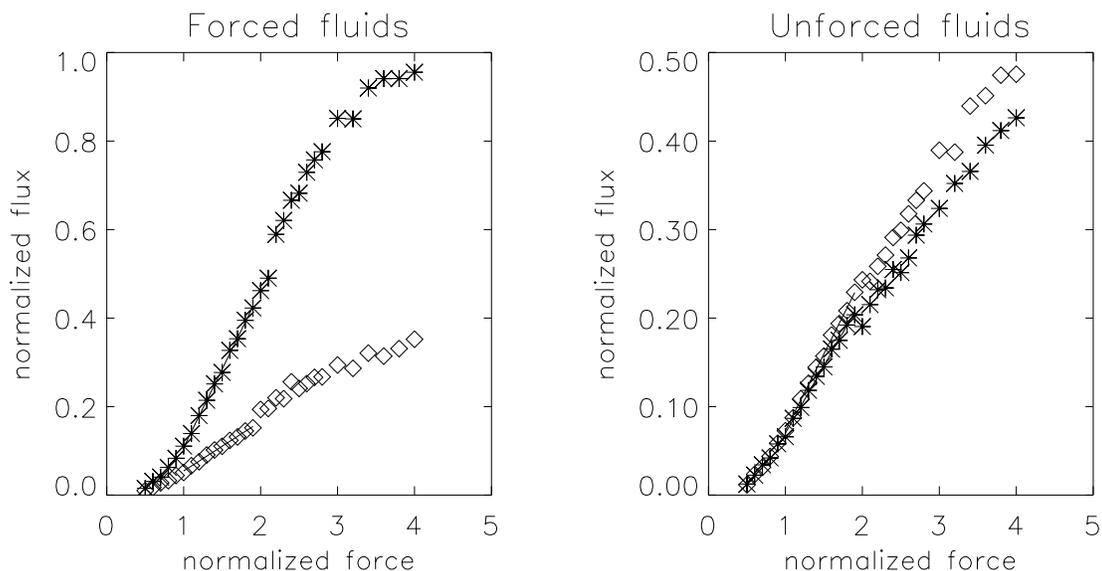}}
\end{center}
\caption{\textbf{Response of fluids when they are either force (left)
or unforced (right).
Stars and diamonds are for oil and water respectively. The normalised force is the
force divided by the force at which linear behaviour first arises. The
normalised flux is the flux divided by the flux of a single component fluid at
that forcing.}}
\label{lijpipe}
\end{figure}

The coupling between the two fluids is studied here. In these
simulations, only one fluid is forced, and the response of both
fluids (forced and unforced) is calculated, for different applied
forces. The results are plotted in figure~\ref{lijpipe}. Simulations
were performed over $20000$ timesteps, with a reduced density of
$0.25$ for each fluid. The wettability of the wall is $-7$, i.e.
strongly water-wetting.\\
\begin{figure}[p]
\begin{center}
\scalebox{1.0}{\includegraphics{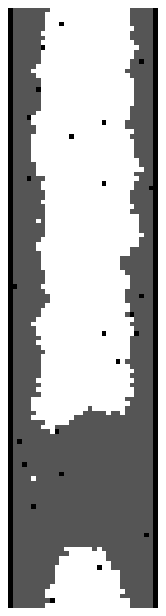}\includegraphics{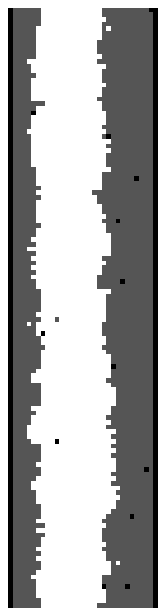}\includegraphics{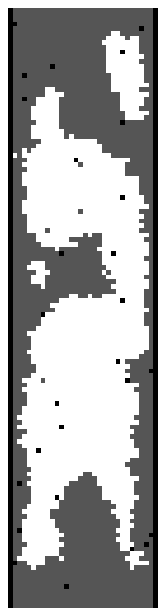}}
\scalebox{1.0}{\includegraphics{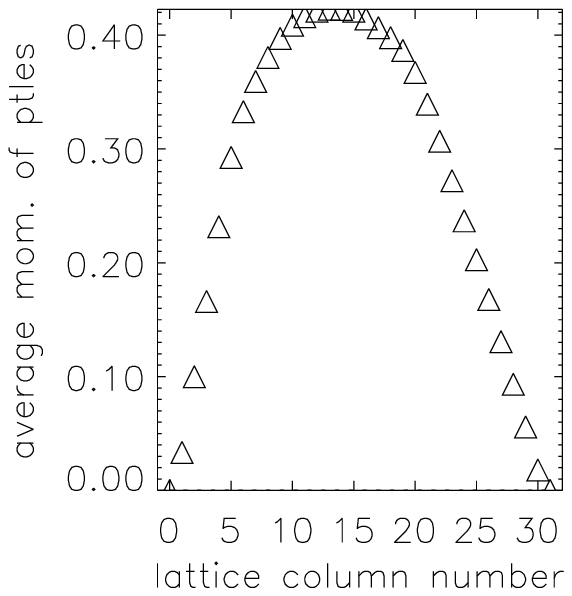}\includegraphics{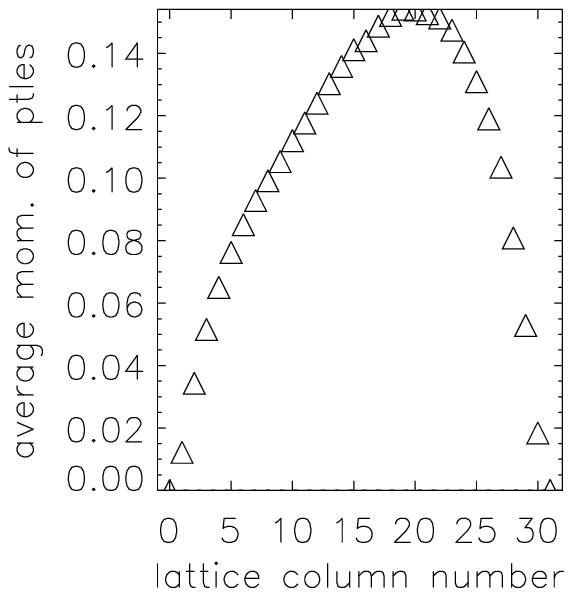}}
\end{center}
\caption{\textbf{A configuration showing the oil phase (white) as a single bubble
(low forcing), a continuous phase (intermediate forcing) or disconnected
phase (high forcing). The velocity profiles, averaged over 10000
timesteps, correspond to a
continuous oil phase when either oil (left) or water (right) is
forced. The lattice size is $32 \times 128$.}}
\label{conflij}
\end{figure}
The behaviour of the two fluids when they are forced is
different. The curve associated with water lies below the one for the
oil because of the wettability of the wall. At small oil forcing levels, the
oil phase exists as a single elongated bubble, and its response is
non-linear. When the force becomes large enough, the response of oil
becomes linear. For an applied force greater than $2$ normalised units, a
discontinuity is seen, which corresponds to the formation of an
infinite lamellar phase (the bubble now extends over the whole
lattice). In this regime, the response of oil is still linear. At
very high forcing levels, the response again becomes non-linear. This is presumably due to
the limitation in fluid velocity of the spatially discrete lattice gas model.\\
By contrast, the response of water is linear from the
smallest forcing levels to a force of approximately $2$ normalised
units (figure~\ref{lijpipe}). The breakpoint
at this forcing level corresponds, as in the case of the forcing of oil, to the
formation of a continuous infinite oil phase. In this regime, the
response is still approximately linear. Increasing the water forcing level
leads to the break up of this infinite oil phase into several oil
droplets, which are no longer ellipsoidal in shape.\\
On the other hand, the response of both fluids is identical when they
are unforced but driven by coupling to the forced second phase:
linear behaviour is observed for a forcing level up to $2$ normalised units. For higher
forces, the response is roughly linear, albeit a little lower than
before, due to the formation of a continuous oil phase.\\
Figure~\ref{conflij} illustrates the different behaviour observed.
On the velocity profile corresponding to a continuous oil phase
when water is forced, a parabolic profile can be discerned on the right
hand portion of the plot (where water occupies the lattice) while a linear
segment on the left hand side corresponds to the location of the  oil,
which behaves as if it were under shear (couette) conditions. When oil is
forced, the central and left hand part of the velocity profile is parabolic
(Poiseuille flow for the central oil phase) while the linear
behaviour is associated with the location of water at each
side of the channel, under conditions approximating couette flow.\\
\begin{figure}[p]
\begin{center}
\scalebox{0.8}{\includegraphics{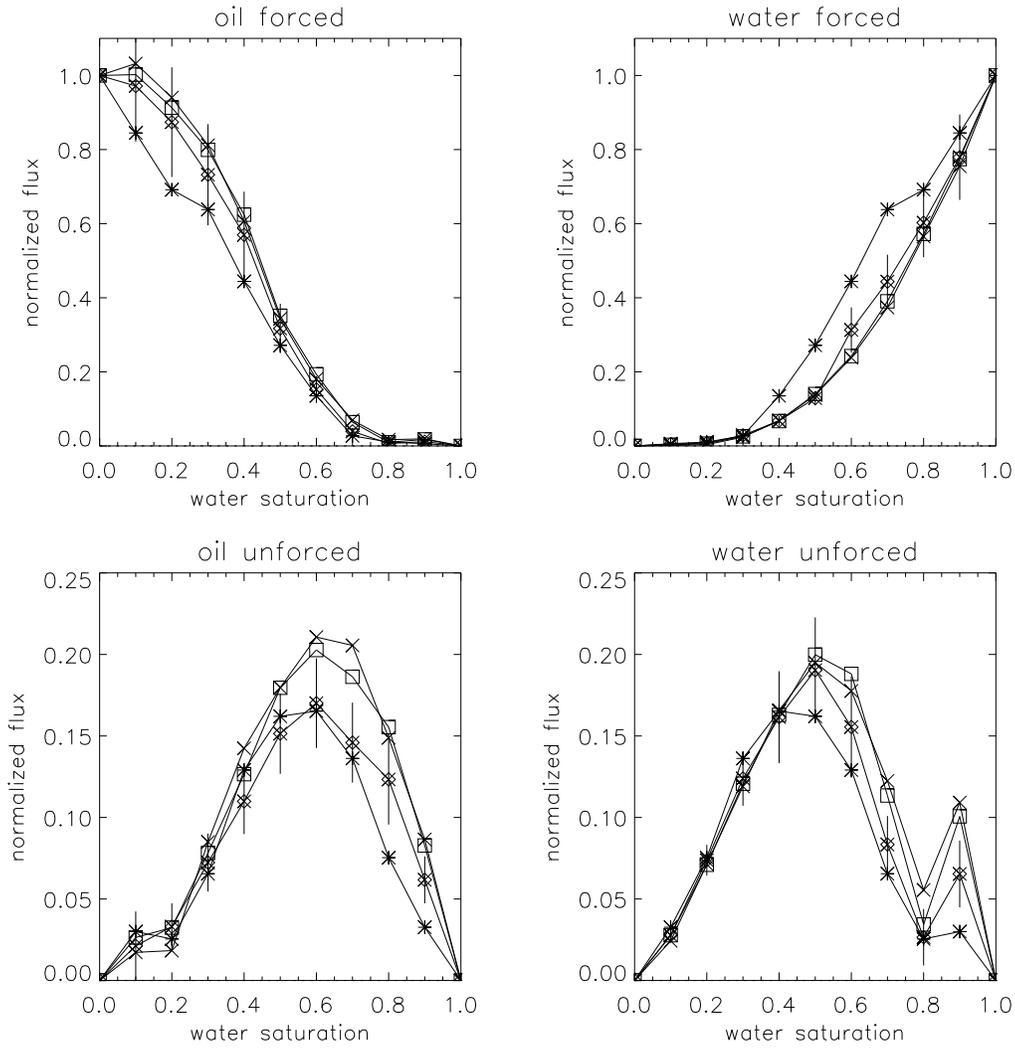}}
\end{center}
\caption{\textbf{Dependence of fluid flow on fluid composition as a
function of wall wettability for progressively less water-wetting
channel walls; the wettability index has the numerical value $-7$
(crosses), $-2$ (squares), $-1$ (diamonds) and $0$
(stars). The normalised momentum is the momentum divided by the
momentum of a pure fluid at the same level of forcing. Data are
averaged over 10000 timesteps on a $32 \times 128$ lattice. For
clarity, error bars are displayed only for one set of simulations.}}
\label{pipelij}
\end{figure}
Figure~\ref{pipelij} displays the results of the calculation of the
relative permeability coefficients over the whole range of water
concentration, and for different channel wettability indices.
In the case where oil is forced, the plots of the oil response
become more curved as the wettability of the wall increases (
the curve is convex up to $0.4-0.5$ water concentration, and concave
for higher concentrations). Moreover,
lubrication effects are seen at small water concentrations, and they
become more important as the wettability coefficient increases
(lubrication is manifested by
a normalised momentum greater than one).
This is due to the fact that when the wettability coefficient is
small (between $0$ and $2$), the water preferentially accumulates at
only one side of the walls. Thus the oil is in direct contact
with the other wall. For the highest wettability index (-7), the
water adheres to both sides, and the oil flows in the centre of
the pipe, avoiding any contact with the walls. Thus the oil
does not dissipate momentum on the wall anymore, and the resulting
flow is greater than that of a pure fluid.\\
By contrast, when water is forced, the concavity of the
flux-composition curves increases with
the wettability index, and the curves remain concave over the
whole composition range.\\
The main effect of increasing the wettability index on the
response of unforced fluids is that the viscous coupling between
the two fluids then also increases. Moreover, a secondary maximum
appears, when the wettability decreases (in the case of unforced oil
at low water concentration) or increases (in the case of unforced water at
high water concentration). In the latter case, this peak is
related to the fact that at very high water concentration, there is a small
(roughly spherical) oil bubble which can flow in the pipe (whose
diameter is lower than the diameter of the pipe). When the water
concentration decreases, the diameter of the oil bubble increases and
becomes identical to the diameter of the pipe, at which point it
experiences direct interaction with the wall.
In the case of unforced oil, the secondary maximum is
associated with the formation of water droplets
rather than elongated layers along each wall.
These droplets can flow easily, and the resulting coupling with
water is enhanced. As in the case of forced fluids, the main effect of the wettability
index appears in the range $0-2$.

\section{Darcy's law and its generalisations}
In this section, we are concerned with the study of fluid flow in
porous media. The first sub-section describes the method used to construct a porous
medium, which differs from the random distribution of obstacle
sites used previously~\cite{coveneymaillet}. In the case of a
single component fluid, the validity of Darcy's law is investigated, for
various porous media. In the case of
binary immiscible fluids, there is no generally accepted law
governing the flow behaviour of the mixture. A generalisation of
Darcy's law (eq~\ref{generaldarcy}) is used in previous work, which explicitly admits
viscous coupling between the two fluids. The importance
of this coupling is examined for various porous media.
The dependence of fluid flow on the fluid composition is
investigated. Finally, an
investigation of the influence of surfactant is also presented.
\subsection{A brief review}
There have been a few studies concerning the calculation of relative
permeabilities of two-phase flow in porous media. In two
dimensions, Rothman~\cite{rothman90} used a lattice-gas
method to investigate the validity of macroscopic fluid flow laws in porous
media. He constructed a porous medium comprised of a square block in the
middle of a pipe. For each saturation, he computed the four
phenomenological coefficients (see equation~\ref{generaldarcy}) from the slope of the linear response of
the flux (of each species) to the applied force. He then constructed
a relative permeability diagram, which shows that the viscous
coupling is not negligible (typically of the order of $0.2$ for a $50\%$
saturation). Nevertheless, his model ``porous medium'' was extremely simple.
Kalaydjian~\cite{kalaydjian90} compared theoretical
predictions with experimental results on the behaviour of an oil
ganglion in a capillary tube square cross section and axial
constriction. He found that for a ratio of viscosities equal to one,
the effect of viscous coupling is significant.
Moreover, taking into account variation of the viscosity ratio of the
two fluids, the relative permeability can assume values greater than one,
indicative of lubrication effects.
Goode and Ramakrishnan~\cite{goode93}, have calculated relative permeabilities
in the case of a tube network using a finite element method. They
found that the viscous coupling is very small.
They also studied the influence of the viscosity ratio on
lubrication. Zarcone \emph{et al.}~\cite{zarcone94} have
constructed an experimental setup to
determine the coupling coefficient from a single experiment. The
porous medium was a packing of sand grains and the experiments were
performed on pairs of fluids (mercury/water and oil/water). They
assumed that the cross-coefficients were equal over the whole range of
saturation and found that they could be neglected in both cases.
In a recent study, Olson and Rothman~\cite{olson97} have calculated
relative permeabilities using the lattice-gas method in three dimensions,
for a digitized microtomographic image of Fontainebleau sandstone. The
coupling coefficients appear to be very
small and Onsager's reciprocity relation is verified.

\subsection{Construction of two-dimensional porous media}
\label{construction}
\begin{figure}[ht]
\begin{center}
\scalebox{1.0}{\includegraphics{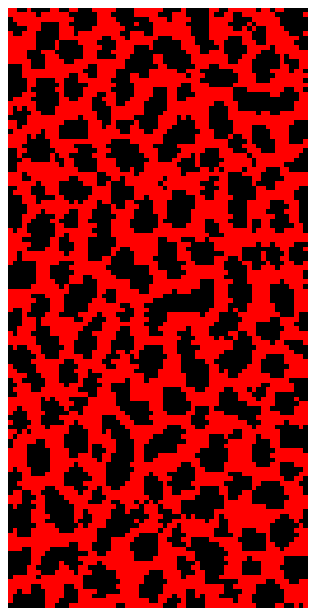}\includegraphics{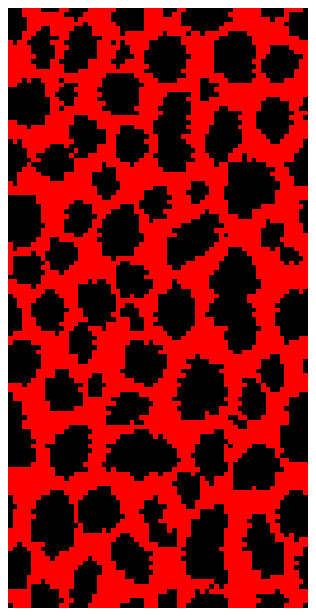}\includegraphics{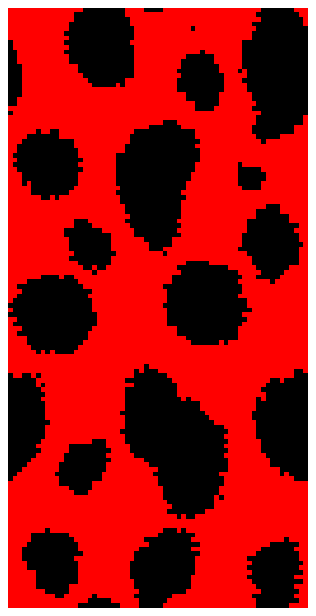}}
\end{center}
\caption{\textbf{Examples of porous media ($64 \times 128$ lattice
sites) constructed from domain growth in a binary
phase. Obstacle sites are coloured in black. (right): small, (center) medium,
(left) large.}}
\label{64_128porousmedia}
\end{figure}

It is known that the  behaviour of fluids in porous media changes
dramatically when going from 2D to 3D and it is frequently stated that
there is no such thing as a porous medium in two dimensions. Nevertheless, 2D simulations can
be compared to 2D-like micromodel experiments, and prove to be helpful
in understanding this complex problem.
The first step needed before doing any simulation is the construction of
a porous medium itself. If we take a 2D slice of a 3D porous medium, we cannot be sure that the pores are connected (the connection may appear in
the third dimension), and thus simulations in such media would be
impractical. On the other hand, a regular porous medium
can introduce unwanted symmetry and thus introduces artifacts into the results.\\
The method used by Wilson and Coveney~\cite{wilsoncoveney} to create a porous medium
consists of randomly placing obstacle sites on the lattice. This has
the disadvantage that the resulting medium has rather uncontrollable properties.
The method proposed here ensures control of the size and
dispersion of solid obstacles without imposing any symmetry: first a
simulation of domain growth in an oil-water binary mixture is run,
starting from a random configuration. The
temporal evolution of such a mixture is marked by the formation of small
droplets of oil in bulk water~\cite{boghosian}.
These droplets coalesce with each other, becoming larger and larger. When the
size of the droplets reaches the desired size for the obstacle, the
simulation is stopped and the oil sites are stored
as obstacle sites for latter use. Figure~\ref{64_128porousmedia} shows
examples of porous media constructed in this way, which have been
extensively used in the simulations described here.\\
One should note that the porosities of these media are
roughly equal to each other (table~\ref{porousmed}); the differences originate from the
average size of the obstacle grains.
The use of these kinds of porous media enables us to readily change
the scale of the simulations.

\begin{table}[ht]
\begin{center}
\caption{\textbf{Properties of the three porous
media (small, medium and large) shown in
figure~\ref{64_128porousmedia}. The effective surface is the number of
available lattice sites
in direct contact with an obstacle site. A site is far from an obstacle
if the distance to the nearest obstacle site is larger than one
lattice site.}}
\label{porousmed}
\vspace{2.0mm}
\begin{tabular}{ccccc}
\textbf{Properties} & \vline & \textbf{Small} & \textbf{Medium} &
\textbf{Large} \\
\hline\hline
Porosity & \vline & $55\%$ & $52\%$ & $60\%$ \\
Number of available sites & \vline & $4466$ & $4230$ & $4929$ \\
Effective surface & \vline & $3288$ & $2407$ & $1101$ \\
Number of sites far from obstacle & \vline & $1178$ & $1823$ & $3828$
\end{tabular}
\end{center}
\end{table}

\subsection{Single phase fluids}
\begin{figure}[ht]
\begin{center}
\scalebox{1.0}{\includegraphics{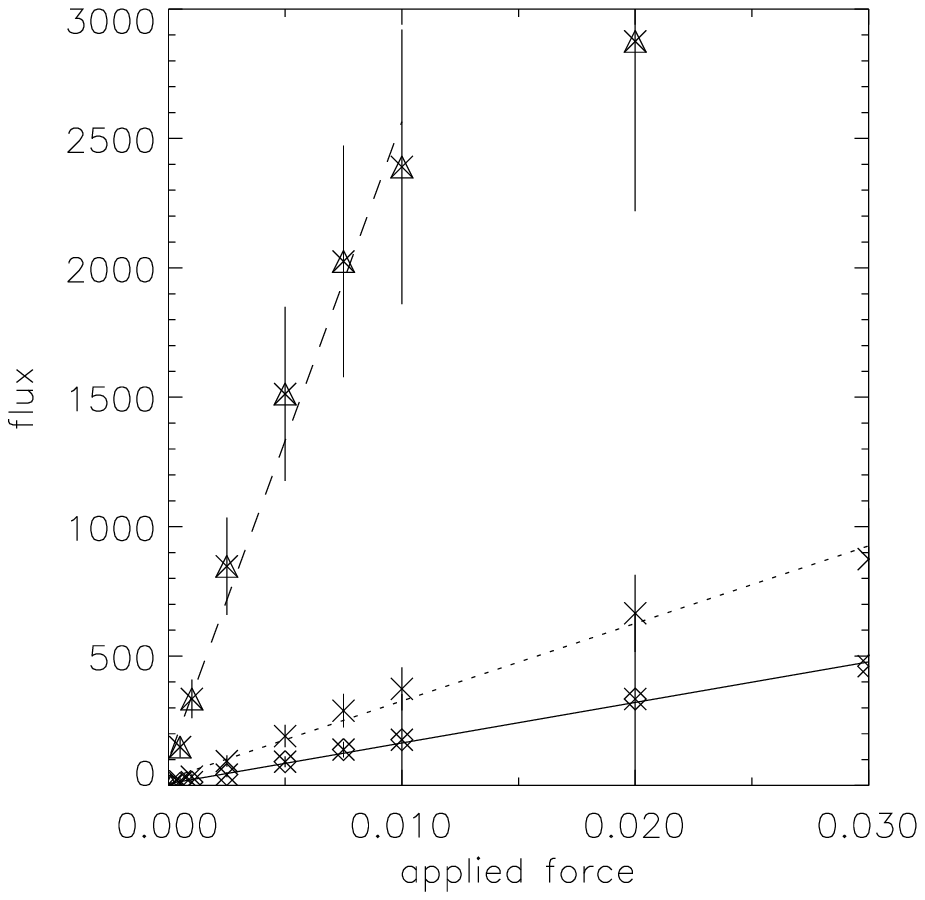}}
\end{center}
\caption{\textbf{Verification of Darcy's law for a single component
fluid in the three different
porous media of table~\ref{porousmed} and \ref{64_128porousmedia}. The
lines are linear fits to
the data. Triangles, crosses and diamonds refer to the large, medium and small
porous media respectively.}}
\label{monophasedarcy}
\end{figure}

In the case of single component fluids, the flow is governed by Darcy's law:
\begin{equation}
\mathbf{J} = -\frac{k}{\mu} (\mathbf{\nabla} p - \rho \mathbf{g})
\end{equation}
where $\mathbf{J}$ is the flux, $k$ the permeability, $\mu$ the
kinematic viscosity, $\mathbf{\nabla} p$ the pressure gradient and $\rho \mathbf{g}$ the
gravitational force density. To determine permeabilities for these
three porous media, simulations were performed at a reduced density of 0.6 particles per
lattice direction with an initially random configuration. The flux was averaged
over the entire lattice and over 7500 timestep intervals, after the
flow had reached a steady state. 
Figure~\ref{monophasedarcy} displays the results calculated over a
wide range of applied forces, making use of the gravity condition.\\
In the three porous media (cf table~\ref{porousmed}), a linear
expression fits the simulated
points: Darcy's law is thus obeyed in these cases. However, for very high
levels of forcing, nonlinear effects appear (for the largest porous medium).
The slope of the linear fit is proportional to the permeability of
the porous medium; the ratios of these slopes are reported in table~\ref{perm}.

\begin{table}[ht]
\begin{center}
\caption{\textbf{Ratios of the permeabilities ($k_s,k_m,k_l$) of the small, medium, and large porous media respectively.}}
\label{perm}
\begin{tabular}{ccccc}
Ratio of permeabilities & \vline & $\frac{k_{m}}{k_{s}}$ &  $\frac{k_{l}}{k_{m}}$ &
$\frac{k_{l}}{k_{s}}$ \\
\hline\hline
Numerical value & \vline & $1.8$ & $8.5$ & $15.7$ 
\end{tabular}
\end{center}
\end{table}
All these simulations were performed under the same conditions:
the viscosity of the fluids is unchanged and the porosities of the various
porous media are very similar to one another. The slopes of the lines, i.e.
the permeabilities of these three porous media are strongly dependent on the
chosen resolution (averaged number of lattice sites per pore).

\subsection{Binary immiscible fluids}
Even binary immiscible fluid flow in pipes is a problem for
conventional continuum fluid dynamics methods, because of the complexity of
the fluid interfaces, particularly at high Reynolds numbers. Similar
challenges remain for the modelling and simulation of two-phase flow
in porous media, even at very low Reynolds numbers.

\subsubsection{Cross-coefficients and Onsager relations}
\begin{figure}[hp]
\begin{center}
\scalebox{0.7}{\includegraphics{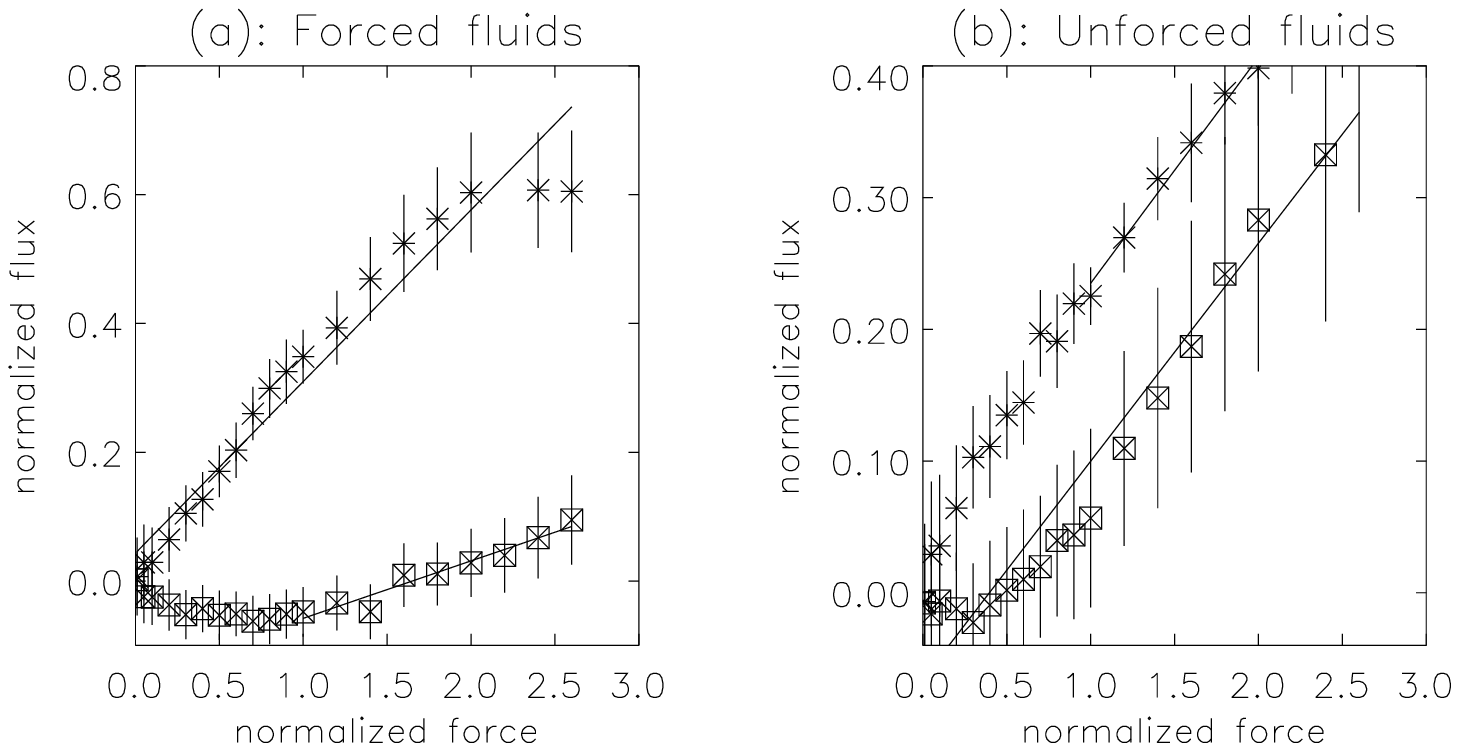}}
\scalebox{0.7}{\includegraphics{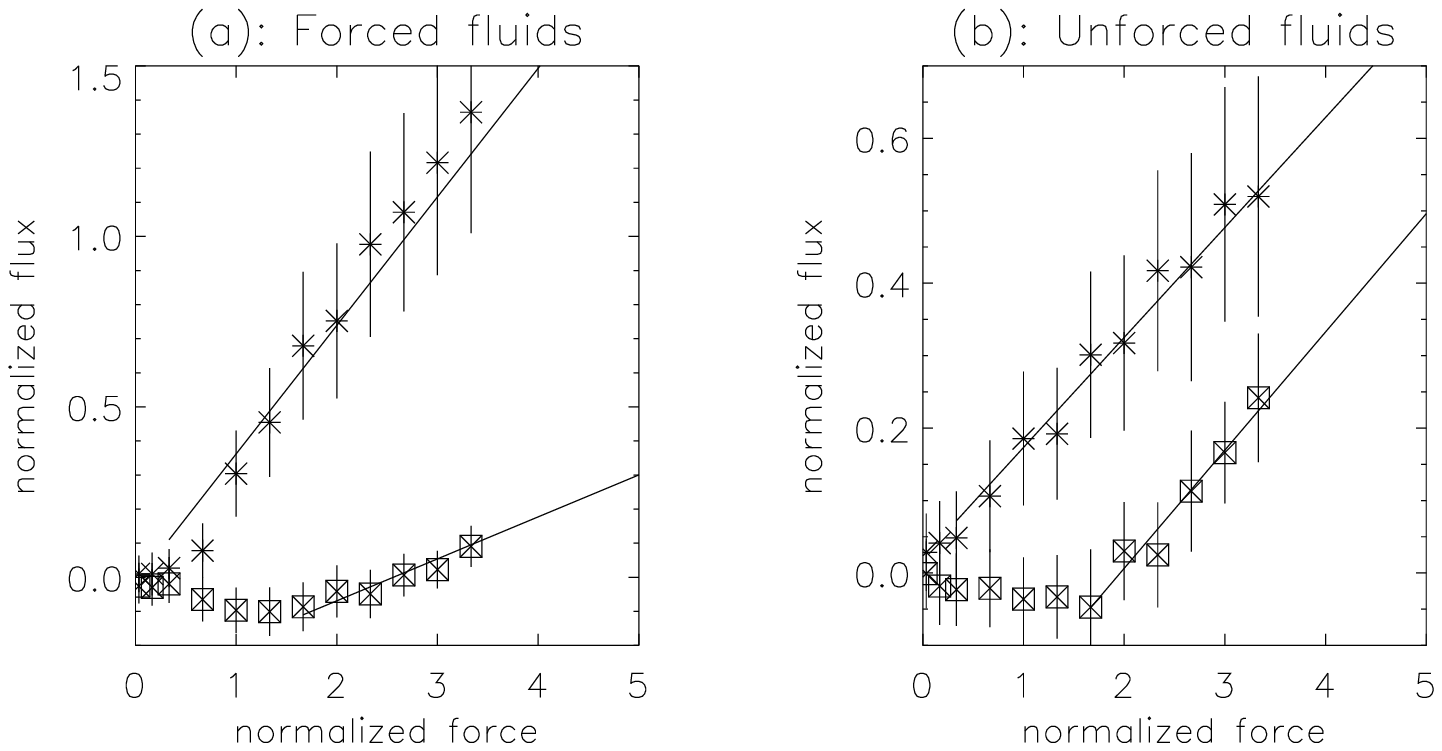}}
\scalebox{0.7}{\includegraphics{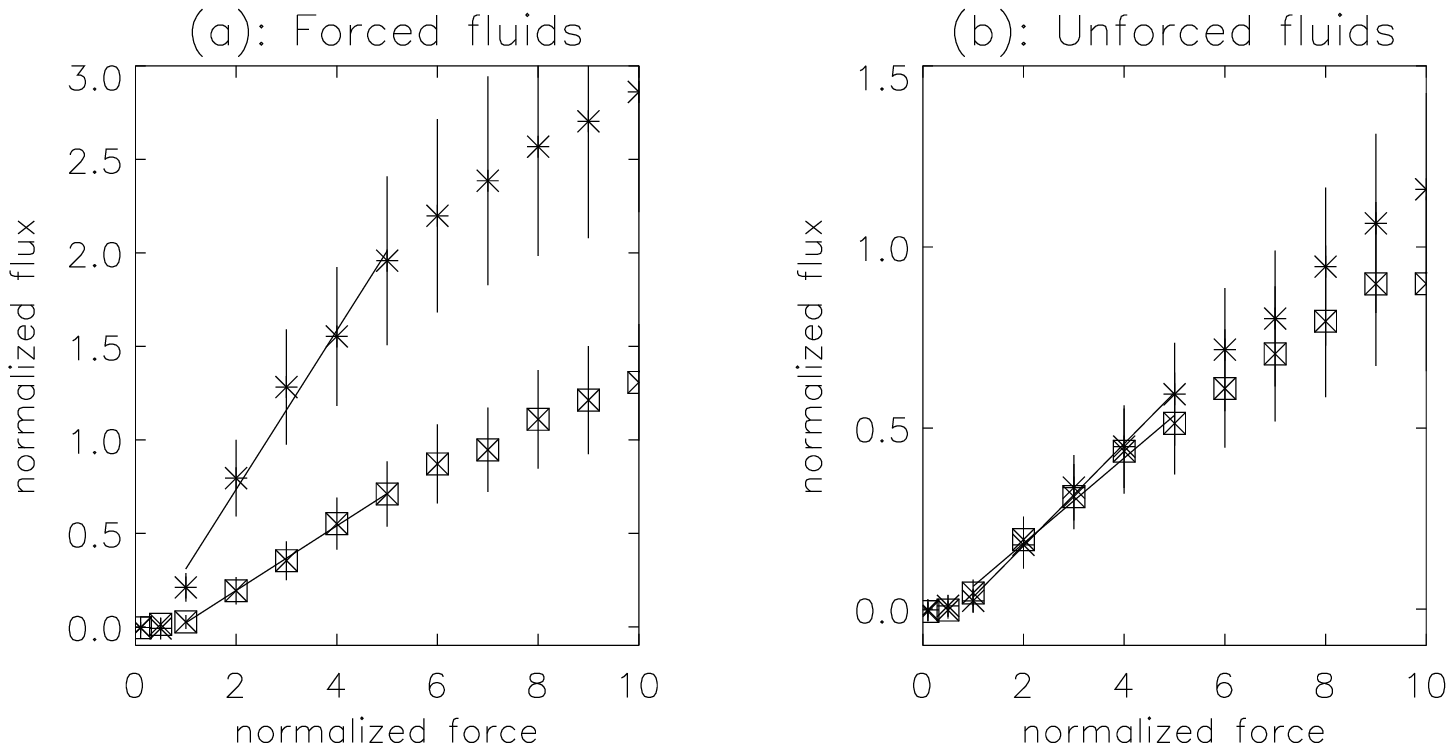}}
\end{center}
\caption{\textbf{Investigation of a generalised version of Darcy's
law: calculation of the phenomenological coefficients as the slope
of the response of each fluid component when they are either forced (diagonal)
or unforced
(cross-coefficients) in the case of the small (top), medium (centre), and large
(bottom) porous media of table \ref{porousmed}. Oil and water are
represented by stars and squares respectively. Data are averaged over
10000 timesteps on a $64 \times 128$ lattice.}}
\label{lijsmall}
\end{figure}

At the outset, we should stress that there is no well-established law
governing the flow of binary immiscible fluids through porous media.
One might consider the simplest extension of Darcy's law for two
components of the form:
\begin{equation}
\mathbf{J_{i}} = k_{i}(S) \frac{k}{\mu_{i}} \mathbf{X_{i}},
\end{equation}
where $\mathbf{J_{i}}$ is the flux of the $i$th species, $k_{i}$ is the
relative permeability coefficient (depending on the saturation $S$),
$\mu_{i}$ is the viscosity, and $\mathbf{X_{i}}$ is the body force
acting on the $i$th component.
These equations assume that the fluids are uncoupled (i.e. each
fluid flows in a porous medium formed by original the porous medium plus the
other fluid).\\
However, if we assume that the fluids are coupled, the
generalised form of Darcy's law becomes:
\begin{equation}
\mathbf{J_{i}}= \sum_{j} {L_{ij} \mathbf{X_{j}}},
\label{generaldarcy}
\end{equation}
where $i,j=1,2$. Eq~\ref{generaldarcy} is a more general form of
linear force-flux relationship. These equations have a similar
structure to that which arises in the theory of linear irreversible
thermodynamics. The
coefficients $L_{ij}$ are then referred as to phenomenological coefficients. In this
theory Onsager's reciprocity relation applies (the cross coefficients are
equal ($L_{12} = L_{21}$)). Tempting as it is, one must nevertheless
be very careful about claiming that there is more than a formal
similarity in this case. Onsager's theory depends on various physical
assumptions, such as detailed balance, which are obviously not satisfied here.\\
Simulations have been performed on a 1:1 mixture of water and oil, at a
total reduced density of 0.5, using the
three porous media described in table~\ref{porousmed}. The wettability
index of the rock was set at $-7$ throughout. The applied force
(gravity condition) varies over the range [0.0001, 0.2].
To calculate the different phenomenological coefficients, we computed
the response of both fluids when each one is
forced (figure~\ref{lijsmall}).
In the literature~\cite{olson97}, the force is normalised with respect to the
capillary threshold (the force needed for an oil bubble to flow). In our
work, particularly using the small obstacle matrix, the flux of
water becomes linear at higher forcing levels than for the oil. Thus
the normalisation is made with respect to the appearance of a linear
response for \emph{both} fluids.\\
The response of forced fluids is linear over a large
applied force range in the case of the small porous medium, and for low
forces in the case of the large one. Thus we can extract the coefficients $L_{11}$ and
$L_{22}$ from the slope of the linear regime. The coefficient relative to
the water is lower than the one for the oil because of the wettability
of the porous medium. The water preferentially adheres to the
obstacles rather than flowing in the channels. However, the two diagonal
coefficients have the same order of magnitude. \\
In the case of oil,
capillary effects can be seen at very low forcing levels. This leads
to non-linear behaviour. This non-linearity vanishes for applied
forces greater than a critical value corresponding to the capillary
threshold.  In the case of water, the non-linearity
disappears when the average size of the channels increases (that is,
for the large porous medium). When the channels are very narrow, a small oil
bubble can easily block the flow of water. Thus, this non-linearity
is directly related to the width of the channel; it may also be influenced by
the connectivity of the obstacle matrix (tests have been made to
check the influence of the bounce-back (no-slip) boundary conditions at obstacles on this
behaviour. It appears that using other types of conditions does not
change these results).\\

The behaviour of the unforced fluids is quite
different: in the case of oil, the response is
linear, even for small applied forces. This is presumably due to the fact
that oil particles reside preferentially far from
obstacles and so a flux of water particles will necessarily induce a
flux of oil particles. In the case of unforced water, its response is linear
only if the applied force is greater than a critical value (in the
small and medium sized porous media). In the low
forcing regime, the response is clearly non-linear and sometimes even
negative. This non-linearity can be attributed to the wettability of
obstacles with respect to water as well as to the fact that oil
bubbles can block pore throats and so prevent the flow of water. \\
However, the cross-coefficients in the generalised form of Darcy's
law (eq.~\ref{generaldarcy}) are nearly
equal to each other (figure~\ref{lijsmall}). The surprising thing is
that the numerical values of the cross coefficients are of the same order
of magnitude as those of the diagonal coefficients
$L_{11}$ or $L_{22}$. In
comparison with the results obtained by Rothman \emph{et
al}~\cite{bookrothman}, this fact may be attributed to the
dimensionality of
the model (we are working in 2D while ~\cite{bookrothman} is
concerned with the 3D case) and to the porous media used.
The relative permeabilities are reported in table~\ref{relperm} for
each porous medium.
\begin{table}[ht]
\begin{center}
\caption{\textbf{Relative permeabilities for the three porous media of
table \ref{porousmed}, and coupling cross-coefficients. Their ratios
are also reported. Indices 1 and 2 refer
to oil and water respectively.}}
\label{relperm}
\begin{tabular}{ccccc}
$\mathbf{Obstacle}$ & \vline & $\mathbf{Small}$ & $\mathbf{Medium}$ &
$\mathbf{Large}$ \\
\hline\hline
$L_{11}$ & \vline & $0.27$ & $0.38$ & $0.42$ \\
$L_{22}$ & \vline & $0.09$ & $0.12$ & $0.17$ \\
$L_{12}$ & \vline & $0.17$ & $0.15$ & $0.14$ \\
$L_{21}$ & \vline & $0.17$ & $0.16$ & $0.12$ \\
$\frac{L_{11}}{L_{22}}$ & \vline & $3.0$ & $3.17$ & $2.47$ \\
$\frac{L_{12}}{L_{21}}$ & \vline & $1.0$ & $0.94$ & $1.17$ \\
$\frac{L_{11}}{L_{12}}$ & \vline & $1.59$ & $2.53$ & $3.0$ \\
$\frac{L_{22}}{L_{21}}$ & \vline & $0.53$ & $0.75$ & $1.42$
\end{tabular}
\end{center}
\end{table}
The increase of the relative permeabilities when going from the small
to the large porous medium can be understood in term of connectivity of space: in the case of the small porous medium, pores are
numerous and very small, and the oil phase is much more disconnected
than in the large obstacle matrix, in which the pores are large,
enabling the oil to form larger droplets. Thus the amount of fluid-fluid interface decreases
when going from the small to the large porous medium. This explains
the small decrease of the cross coefficients and the
increase of the relative permeabilities of each fluid.\\
Note that the ratio of cross terms is approximately constant, and equal to
one. This supports the notion of Onsager reciprocity, which seems to be
valid in spite of the fact that various properties, such as detailed
balance, are not maintained by the lattice gas model. Moreover, the
ratio of diagonal terms is also approximately
constant, implying that a correlation exists between the forcing of
oil and water. Nevertheless, the
ratio of diagonal to off-diagonal terms
for each fluid is not constant: this is due to the fact that the
coupling between the two fluids also depends on the geometry of the porous medium.\\
These simulations demonstrate that there is significant coupling
between the two fluids, at least in the case of these particular porous media.
This strong viscous coupling is thought to be due to the dimensionality of the
model. The reciprocity relations seem to be valid over a well-defined
range of applied forces, and the response of the fluids remains linear
down to a well defined value of the applied force.

\subsubsection{Relative permeabilities as a function of saturation}
\begin{figure}[ht]
\begin{center}
\scalebox{1.0}{\includegraphics{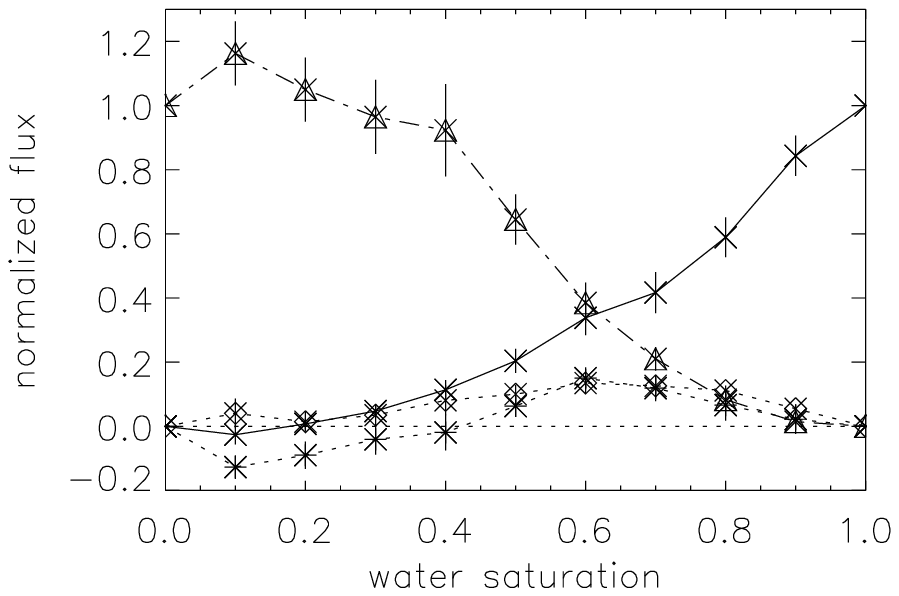}\includegraphics{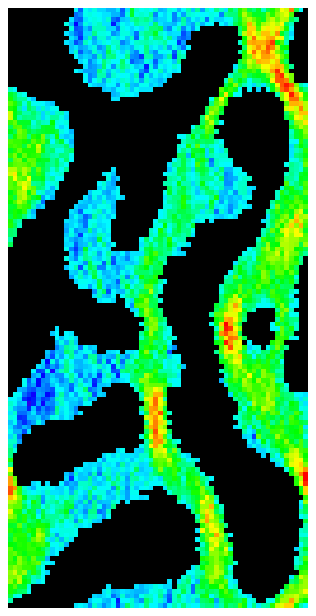}}
\end{center}
\caption{\textbf{Left: binary immiscible fluid flow behaviour as a function
of water saturation. The normalised momentum is the total vertical momentum
divided by the total vertical momentum of pure fluid at the same
forcing level in the same porous medium. Water is represented with crosses
(forced) and stars (unforced) and oil with triangles (forced) and
squares (unforced). Right: coloured velocity profile of
the flow of a single phase fluid in the same porous medium. Red and
blue are associated with large and low velocities respectively. The size of the
porous medium is $64 \times 128$. Averaging is performed over 10000 timesteps.}}
\label{conclam3}
\end{figure}

\begin{figure}[ht]
\begin{center}
\scalebox{1.0}{\includegraphics{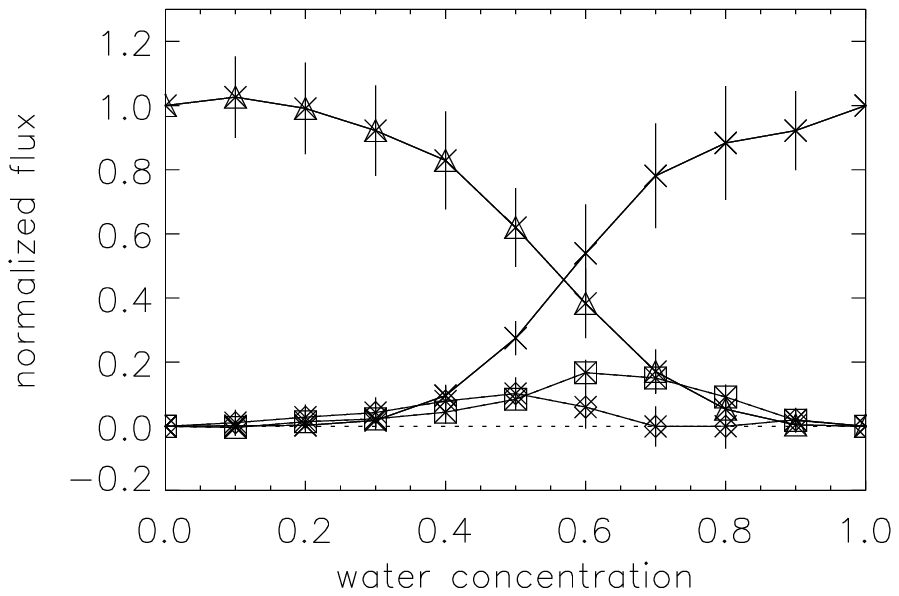}\includegraphics{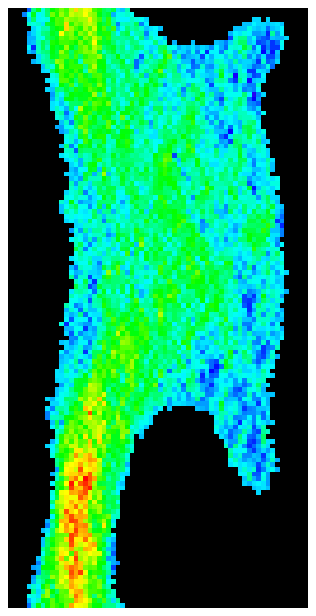}}
\end{center}
\caption{\textbf{Left: binary immiscible fluid flow behaviour as a function
of water concentration. No lubrication phenomenon is exhibited in this
case. Water is represented with crosses (forced) and squares
(unforced) and oil with triangles (forced) and diamonds (unforced).
Error bars are also displayed.
Right: the single pore porous medium used, together with a
velocity profile; colour coding as in previous figures. }}
\label{concha}
\end{figure}

In the last section, we calculated the cross-coefficients in a
linear generalisation of Darcy's law in various porous media. These
results were obtained by looking at the response of each fluid when they are either
forced or unforced, for different applied forces. We focus now on the
behaviour of fluids for different water saturations.
For this study, a new porous medium was selected (see
figure~\ref{conclam3}), because of its large connectivity.
For each saturation, the flux of each fluid is calculated when it is forced or
unforced. The results are shown in figure~\ref{conclam3}.
These curves have been calculated taking into account only the flux of
majority colour sites, but the results are identical if the total flux of
particles is considered. The wettability of the porous medium
has been set equal to $-7$. The total reduced density is
0.7 and the forcing level is 0.005 (gravity condition). Also shown in
figure~\ref{conclam3} is
the coloured velocity profile in the porous medium considered.
The behaviour of the two immiscible fluids when only water is forced
is described first.
At low water saturations, water particles form small bulk water
clusters together with some isolated particles which adhere to the obstacle matrix. Some of the clusters flow through the channels while others
are trapped in stagnant zero-velocity regions.
When the water concentration increases, the cluster size grows, and the
water still occupies both channel sites and stagnant regions. As the water
concentration increases, there is no change in the sites occupied by
water: the mechanism of flow is unchanged.\\
Consider now the behaviour of the oil
when only water is forced. The coupling is maximum for $60\%$
water, corresponding to a significant water flux. When the water concentration
increases, the oil flux decreases because there is less oil in the medium.
Equally, when the water concentration decreases, the oil flux
decreases because there is not enough water to drive oil. However,
the case of forced oil is different.
At low water saturation, the water still forms clusters
as well as surrounding obstacles but now the clusters are all trapped in
stagnant
zero-velocity zones. The normalised flux of water is zero while
the normalised flux of oil is greater than one. This means that a mixture of oil
together with a small amount of water produces a larger flow than that
of pure oil. This can be explained on the basis of a \emph{lubrication
phenomenon}, as in the case of a flowing binary immiscible fluid in a
pipe (see section~\ref{binaryflow}). As shown
below, this is largely due to geometrical properties of the porous medium.
When the water concentration increases, the oil still flows freely in the
channels. Water clusters accumulate preferentially in stagnant
zones.
When the water concentration increases, the
flux of the unforced water increases, for the same reasons as
before. However, the decrease of the oil flux is associated with a reduction of the
connectivity of the oil phase. At higher water concentrations, the flux
of water decreases because of the decrease of the oil flux.
The very small residual oil saturation is due to the use of random initial
configurations: the results are different in invasion simulations
(sections 5 and 6).\\
\begin{figure}[p]
\begin{center}
\scalebox{1.0}{\includegraphics{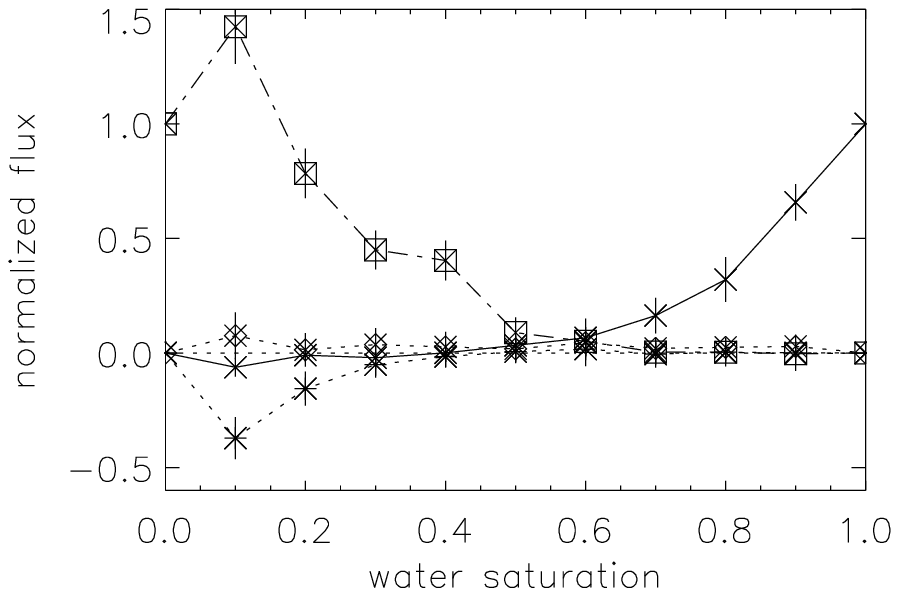}}
\scalebox{1.0}{\includegraphics{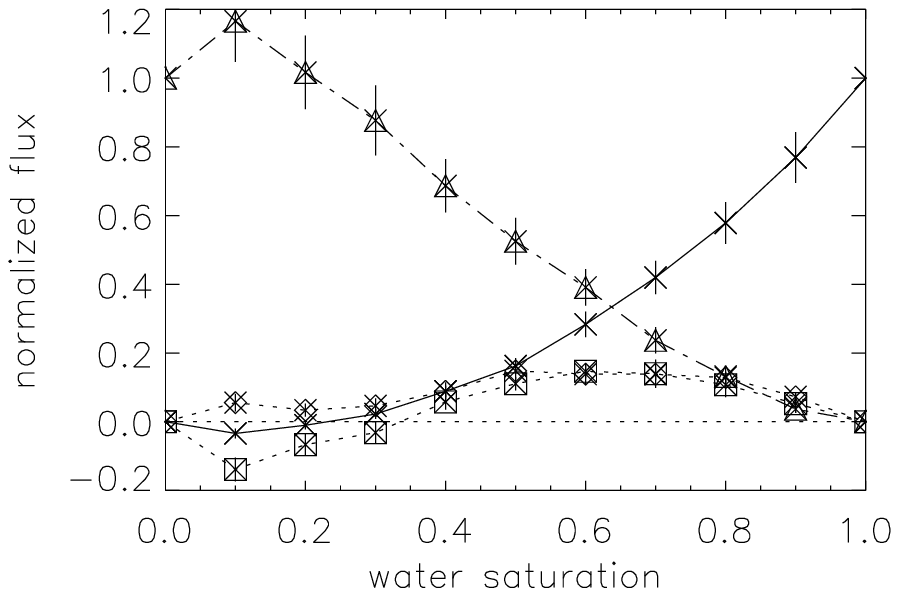}} 
\end{center}
\caption{\textbf{Binary immiscible fluid flow behaviour as a function
of water concentration. The normalised flux is the total vertical momentum
divided by the total vertical momentum of pure fluid at the same
forcing level in the same medium. Water is represented with crosses (forced)
and stars (unforced) and oil with squares (forced) and diamonds
(unforced). The lower graph corresponds
to a forcing level of $0.005$ while the upper one corresponds to
a forcing of $0.001$.}}
\label{conchuge}
\end{figure}
\noindent The same calculations have been performed using the large porous
medium described in table \ref{porousmed}. The results are displayed
in figure~\ref{conchuge} (total density $0.7$, wettability coefficient
$-7$, forcing levels $0.001$ and $0.005$, averaged over $10000$ timesteps).
The two diagrams in figure~\ref{conchuge} represent simulations in the
same porous medium, but with
forcing either greater or lower than the capillary threshold
respectively. In the former
case, the residual oil saturation is very small, and
the effect of the connectivity of the oil phase is small. The
cross-coefficients are roughly equal over a large range of fluid
compositions and the
viscous coupling between the two fluids is non-negligible. The role of
the residual
water concentration is significant, owing to the wettability of
the rock and the connectivity of the obstacle matrix.\\
In the low forcing case (figure~\ref{conchuge}), the residual oil
saturation is very high.
The appearance of a non-zero oil flux is related to the connectivity
of the oil phase: when oil exists only as bubbles, it cannot flow.
The coupling between the two fluids is very small, because of the
high residual saturations (the flux of the forced phase is zero,
as is the flux of the unforced phase).\\
In both diagrams, lubrication phenomena are seen at very low
water concentrations.
Figure~\ref{concha} displays the results obtained in the case of a
porous medium comprising a single pore.
The principal result is that lubrication effects are not very important in this case.
It therefore seems that lubrication effects in the other porous medium used are
due to the topology of the obstacle matrix, being dependent on the
amount of solid-liquid interface.\\
In summary, for binary immiscible fluid flow in porous media, it
appears that when the forcing is greater than the
capillary threshold, the shapes of the relative permeability curves agree well
with what has been published previously concerning simulation results. The
residual oil saturation is
small, and is probably due to the topology of the porous medium and
to the use of random initial configurations.
The porous medium can also produce ``lubrication'' effects, at very
small levels of water saturation. The coupling coefficients appear to be roughly
equal over a large range of saturations, and exhibit a maximum of $0.2$
for $60\%$ water saturation. When the forcing is lower than the
capillary threshold, the oil flux is strongly dependent on the connectivity
of the oil phase.

\subsection{Ternary amphiphilic fluids}
\begin{figure}[ht]
\begin{center}
\scalebox{1.0}{\includegraphics{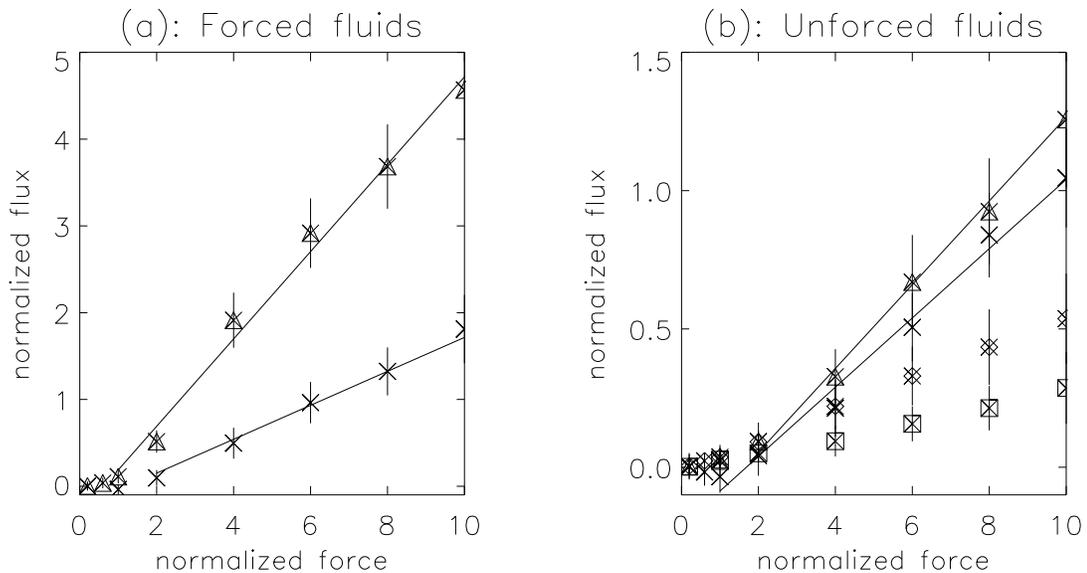}}
\end{center}
\caption{\textbf{Evolution of the response of fluids when they are
either forced (left) or unforced (right), as a function of the forcing level, in the
ternary amphiphilic case, and using the large porous medium (cf table
\ref{porousmed}). Water is represented with crosses (forced and
unforced), oil with triangles (forced and unforced), surfactant with
diamonds (oil forced) and squares (water forced).}}
\label{lijterhuge}
\end{figure}

In this section the effect of surfactant on the
hydrodynamic behaviour of oil/water fluid mixtures in porous media
is investigated.
In the binary immiscible fluid case, the fluxes of the different species were normalised
by the flux of a pure fluid in the same medium, of density equal to
the sum of the partial densities. In the ternary case, the flux of
each species is normalised by the flux of a pure fluid of density
equal to the sum of the partial densities of oil and water.\\
The simulations are performed using the large (cf table \ref{porousmed}) obstacle matrix. As in the
binary case, the gravitational flow implementation is used. The reduced densities of
oil and water are equal to $0.2$, and the reduced density for
surfactant is set equal to $0.1$ (for this surfactant concentration,
the equilibrium state without flow corresponds to the microemulsion phase).
Results are displayed in figure~\ref{lijterhuge}.\\
As in the binary case, the response of fluids when they are forced is
approximately linear, provided the applied force is greater than a
threshold value, the capillary threshold. The diagonal coefficients
(relative permeabilities) can then be calculated from the slopes of these
lines. Their values are ${\it \kappa_{ww}}\approx 0.20$ and ${\it
\kappa_{nn}}\approx 0.50$.
The capillary threshold is lowered by a factor of $\approx 2$
(now equal to $0.0005$) in comparison with the binary case. This
change is due to a lowering in surface tension (oil can now pass more
easily through narrow channels).

\section{Imbibition simulations}
\label{imbibition}
This work follows some initial studies done by Fowler and Coveney~\cite{fower}
on the invasion process in a porous medium. They used a randomly
constructed porous medium made as described in~\cite{wilsoncoveney} to
study the effect of surfactant on the
invasion process in the cases of drainage and imbibition. We have used the same type
of porous media. In this section, we consider only the case of imbibition.
The imbibition process refers to the invasion of a wetting
fluid with or without surfactant into a porous medium filled with a non-wetting
fluid. The aim of the following study is to characterise the invasion
process, and to investigate the
influence of surfactant on this process.
\subsection{Invasion process}
\begin{figure}[ht]
\begin{center}
\scalebox{1.1}{\includegraphics{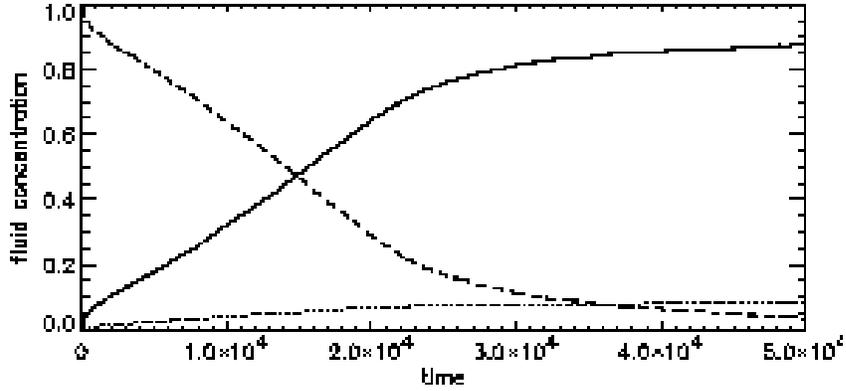}}
\end{center}
\caption{\textbf{Variation of the number of oil (dashed line), water
(continuous line) and
surfactant (dot line) particles \emph{versus} time when a water-wetting
porous medium filled
with oil is invaded by a mixture of water and surfactant. The porous
medium is displayed in figure~\ref{micell}, and the lattice size is
$128 \times 256$. Data come from a single run.}}
\label{fluxinv}
\end{figure}

\begin{figure}[ht]
\begin{center}
\scalebox{0.8}{\includegraphics{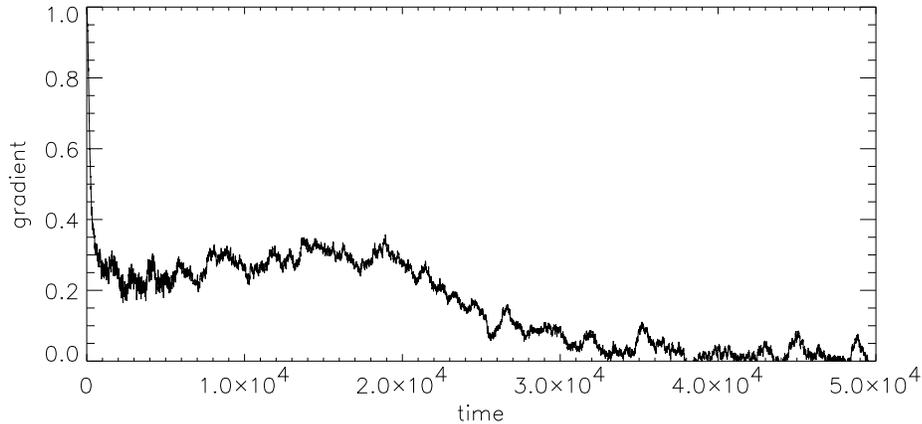}}
\end{center}
\caption{\textbf{Rate of change of water particles \emph{versus} time
(i.e. the water mass flux) for the simulation data shown in figure~\ref{fluxinv}.}}
\label{gradflux}
\end{figure}

A typical result obtained from an invasion simulation, given in terms
of the evolution of the
number of particles of each species versus time, is displayed
in figure~\ref{fluxinv}.
This time evolution can
be decomposed into two parts. In the first regime, at early times in
the simulation, the change in the number of particles is roughly
linear with time. The end of this regime corresponds to
percolation of the water phase, that is breakthrough by the invading
water phase.
The second regime shows a
slow variation of the number of particles of a given type with time, and a
continuously connected path of the invading fluid exists across the
lattice. In this domain a steady state has been reached.
The number of particles of a given type tends slowly to an asymptotic
value. This asymptote
corresponds to the residual oil saturation.
Figure~\ref{gradflux} displays the gradient of the number of water
particles \emph{versus} time, or the water mass flux, for the simulation data shown in
figure~\ref{fluxinv}. The
two regimes described above can be identified here more clearly.
Prior to water percolation, the decrease in the number of oil particles is
linear and its gradient is constant in time. 
After this period, the gradient assumes a small value, corresponding to the
existence of a continuously connected pathway of the invading fluid
across the lattice simulation cell.\\
Between the two domains described above, there is a transient region,
in which the gradient decreases from its constant finite value to
zero. At the end of the first time interval, the water attains its percolation
threshold. At the beginning of the second temporal
domain, the flux of oil is very small and the oil forms only
disconnected bubbles. In the transient regime between these two domains, both
fluids are percolating and the flux of each species is non-zero. Oil
particles are driven off the lattice while water occupies more and more
sites.
The end of the transient domain is associated with the end of oil
percolation.
\subsection{Effect of forcing}
\begin{figure}[ht]
\begin{center}
\scalebox{1.5}{\includegraphics{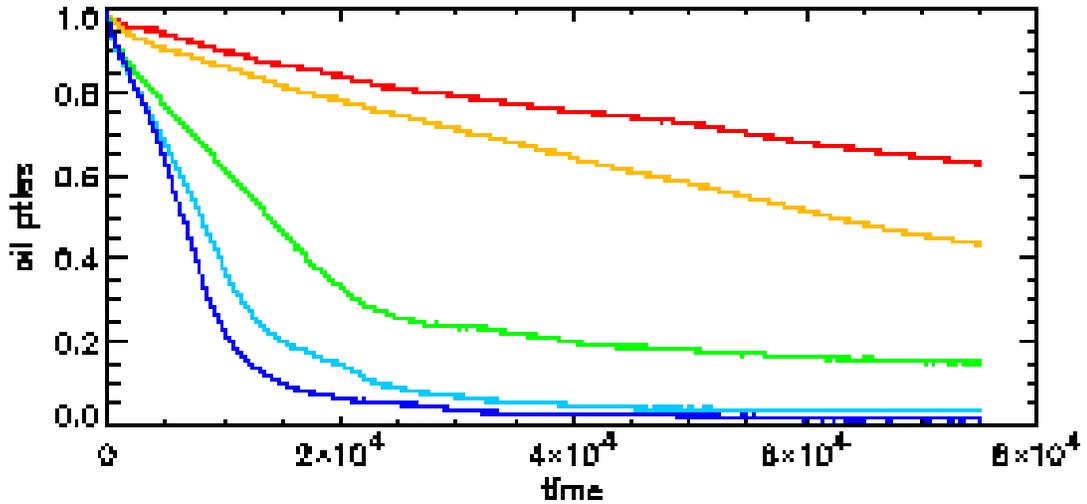}}
\end{center}
\caption{\textbf{Invasion of water in a water-wetting porous
medium filled with oil (imbibition), showing the effect of the applied
force on the flux of oil particles. The time is the number of
timesteps and
the y-axis is the normalised number of oil particles in the lattice. The porous
medium has dimensions $64 \times 128$. Each curve is a result of a
single simulation.
Before percolation, the flux exhibits a linear dependence on time.
The different forcing levels (from top to bottom) are 0.0005, 0.001,
0.005, 0.01, 0.025 (using gravity conditions).}}
\label{forceoncurve}
\end{figure}

\begin{figure}[ht]
\begin{center}
\scalebox{1.0}{\includegraphics{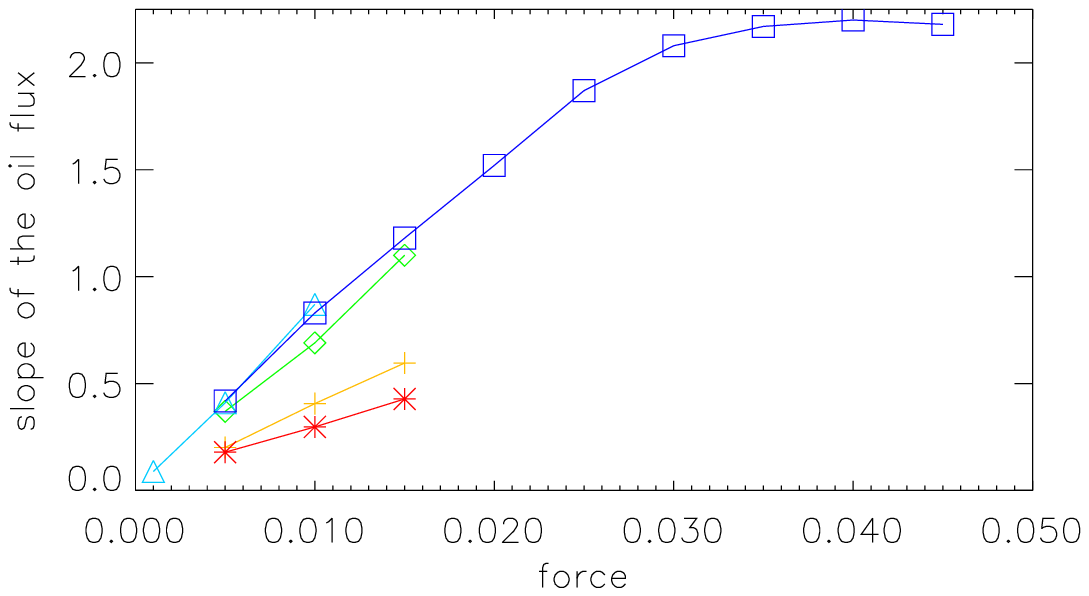}}
\end{center}
\caption{\textbf{Effect of the applied force on the gradient of the flux of
oil particles for porous media (randomly created and also
built using the method described in section~\ref{construction}) of
various size ($64 \times 128$ and $128 \times 256$). Each point is a
result of a single simulation.}}
\label{forceonslope}
\end{figure}

The effect of the applied force is investigated in the first steps of
the invasion process, that is prior to the onset of water percolation.
In this regime, the number of oil particles decreases
linearly with time (the oil mass flux is essentially constant).
Figure~\ref{forceoncurve}
displays the results obtained for the small porous medium (cf table \ref{porousmed}), with no
surfactant, for different applied forces. The simulations were
performed over 75000 timesteps.\\
A linear variation of the number of oil particles versus
time before water percolation is observed. The slopes of these lines
are calculated
for different applied forces. This calculation was performed for
several different porous media, and the results are collected in
figure~\ref{forceonslope}.\\
From these curves (calculated before the percolation threshold), we
can see that there is a linear regime in the
evolution of the oil flux. The oil flux
is then directly proportional to the
force applied to the water particles. Thus the greater the forcing on water, the
faster the oil flows (and the sooner the percolation threshold is reached).\\
At very high forcing levels, the gradient of the oil flux no longer
exhibits a linear dependence on the applied force but rather tends to
an asymptotic limit. This behaviour is probably related to the limited
velocity of the particles in the lattice gas model; otherwise, it would
mean that an optimal value of the forcing level exists to maximise oil
production, implying that the application of ever increasing forces will not be
beneficial.\\

\subsection{Effect of surfactant on imbibition}
\begin{figure}[ht]
\begin{center}
\scalebox{1.0}{\includegraphics{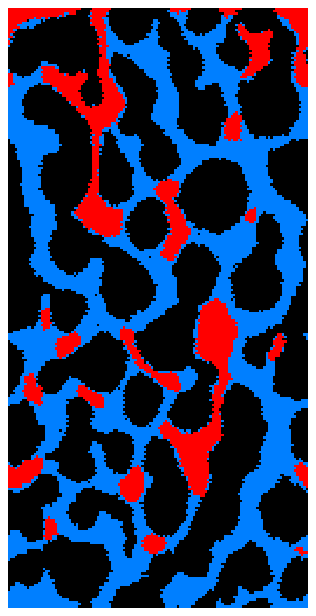}\includegraphics{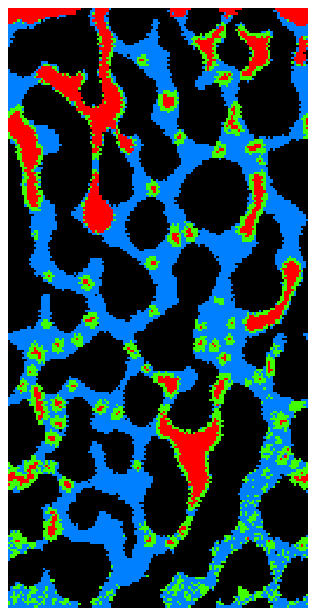}}
\end{center}
\caption{\textbf{Configuration obtained during flooding a porous medium filled
with oil by either water (left) or a mixture of water plus surfactant
(right). The size of the lattice is $64 \times 128$. The
configurations are taken just before the end of oil percolation.
The emulsification phenomenon can easily be seen in the latter case.}}
\label{emulscon4}
\end{figure}

\begin{figure}[p]
\begin{center}
\rotatebox{270}{\includegraphics{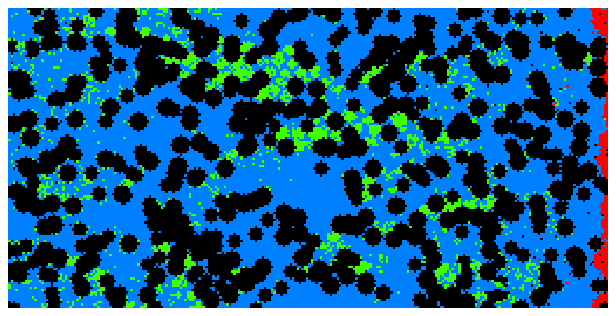}}
\scalebox{0.8}{\includegraphics{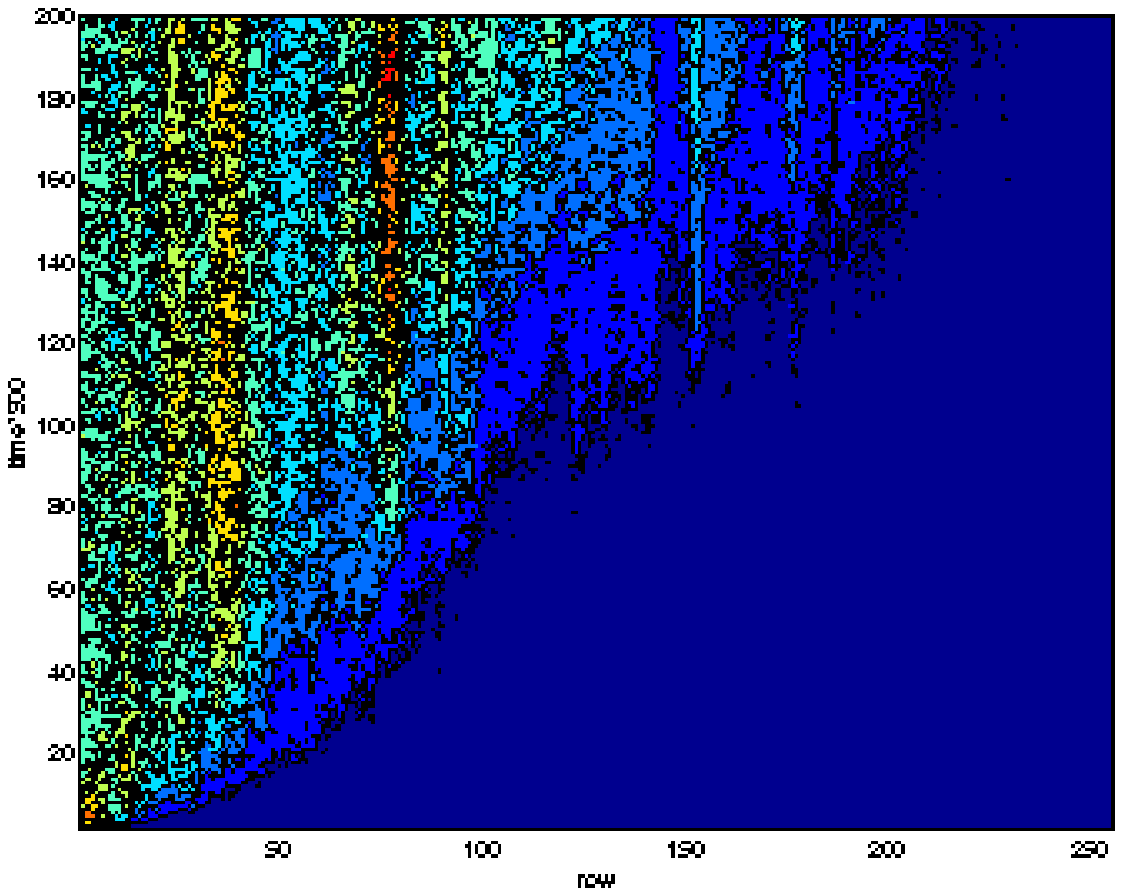}}
\end{center}
\caption{\textbf{Top: configuration after 100 000 timesteps when
flooding a water-wetting porous
medium filled with water by a mixture of water and surfactant (water
in blue and surfactant in green). The direction of the invasion is
from left to right. Bottom: normalised concentration of surfactant \emph{versus} time and
versus the y-coordinate that is, in the direction of the flow (red and
blue correspond to high and low concentration respectively). The
concentration of surfactant in the invading fluid is 0.3. The lattice
size is $128 \times 256$. One time unit corresponds to 250 timesteps.}}
\label{micell}
\end{figure}

\begin{figure}[p]
\begin{center}
\scalebox{0.8}{\includegraphics{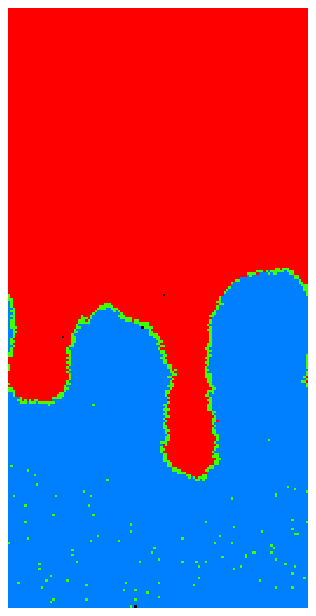}\includegraphics{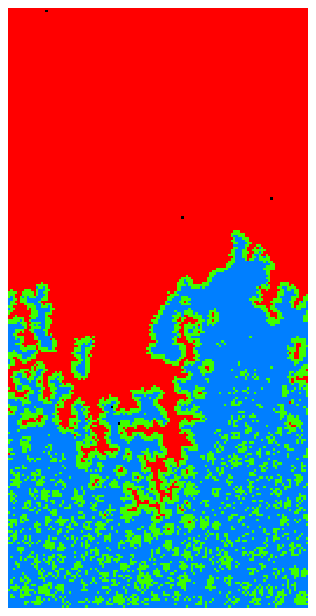}}
\end{center}
\begin{center}
\scalebox{0.5}{\includegraphics{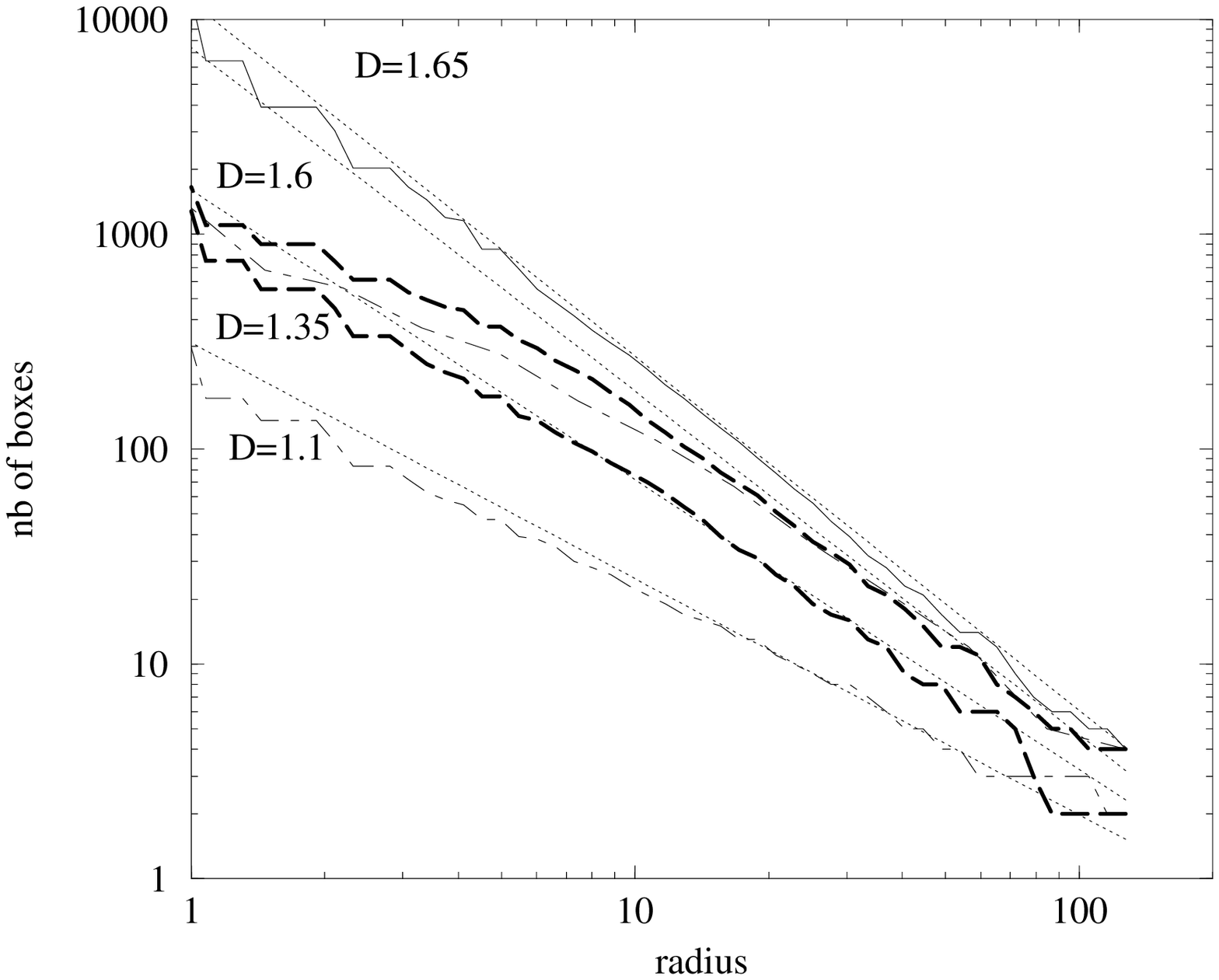}}
\end{center}
\caption{\textbf{Top: configurations during invasion without porous medium, for
concentration of surfactant of $0.1$ (left) and $0.3$ (right). Bottom:
computation of the fractal dimension of the interface (on a log-log diagram
, with a basis of 10).
The lower two curves show the calculation of the
associated fractal dimension in the absence of porous medium
(dot-dashed line for $10\%$
surfactant, dashed line for $30\%$ surfactant). The
curves in the middle ($D=1.6$) correspond to the invasion of binary
(dot-dashed line) and ternary (dashed line) mixture
($\rho_{surf}=0.2$) into one water-wetting (index of
wettability equal to $-7$) porous medium shown in
figure~\ref{micell}. The upper curve (continuous line) is a calculation of
the fractal dimension
associated with the porous medium alone. The y-axis refers to the
number of square boxes (length along the x-axis, in lattice site units) needed to cover the
oil/water and oil/surfactant interfaces.}}
\label{withoutpore}
\end{figure}

\begin{figure}[ht]
\begin{center}
\scalebox{0.5}{\includegraphics{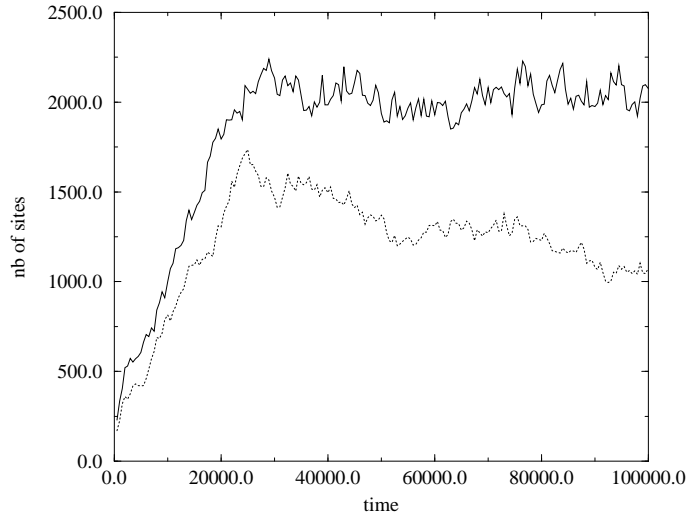}}
\end{center}
\caption{\textbf{Length of the oil-water interface in the binary case
(dotted line) and in the
ternary case ($\rho_{surf}=0.2$)(continuous line), calculated during
invasion into a porous medium (the same as in fig~\ref{micell}, with a wettability index = -7). The
vertical axis measures the number of lattice sites occupied by the
interface and the horizontal axis shows the timestep in the
simulation. The lattice size is $128 \times 256$.}}
\label{lenghtinter}
\end{figure}

\paragraph{Emulsification}
As Fowler and Coveney have noticed~\cite{fower}, one often encounters substantial
variation in results for similar simulations. This means
that precise results require
substantial ensemble averaging, and so one must be very careful in
attempting to draw conclusions from one or a limited number of
simulations. Bearing this in mind, we start by a careful study of the effect of surfactant
at the pore level. The porous medium is now composed of only one
pore as shown in figure~\ref{concha}. This is obviously not
representative of a porous medium but
nevertheless it contributes to our understanding on what happens on
larger scales.
The conclusion arising from this particular porous medium is that the
termination of oil percolation takes longer to occur in the case with
surfactant, and hence that oil recovery is enhanced. The behaviour of
the fluids
after the end of oil percolation is strongly dependent on whether
surfactant is absent or present. Without surfactant, the remaining oil
forms large
spherical droplets which can barely flow, whereas the presence of
surfactant favours the formation of
smaller oil droplets (the surfactant concentration required to achieve this is
quite high, typically $0.2 - 0.3$), which can flow very easily. Therefore,
in this example of imbibition, the introduction of surfactant strongly
enhances oil recovery.\\
These phenomena observed at the pore level are also manifested at larger
scales. Figure~\ref{emulscon4} displays configurations obtained when
flooding a porous medium filled with oil by either water or a mixture
of water and surfactant.
From this figure, we can see that when there is no
surfactant, rather large oil droplets remain in the rock. On the
other hand, in the presence of surfactant, a large number of small oil
droplets can be seen surrounded by surfactant. These small droplets are
less affected by capillary effects and can flow readily. However, we also
notice that the concentration of surfactant is greater in the lower part
of the box, from where the flooding occurred. There are two reasons for
this: first, surfactant particles are rapidly trapped when they
encounter the first oil-water interface. This trapping remains until
the oil droplets form and flow. Second, surfactant particles can
self-assemble into small clusters~\cite{mailletlachet, coveney96}, which can be trapped in small pores.
This phenomenon, called \emph{micellisation}, is discussed further below.\\
These various emulsification phenomena are clearly observed in our imbibition
simulations, and tend to decrease the residual oil
saturation in the steady state regime, thus improving
the oil recovery.
\paragraph{Micellisation}
One important effect arising from the presence of
surfactant is now discussed in greater detail: micellisation. This
corresponds to the
formation of small clusters of surfactant, trapped in the medium. It
requires a sufficiently high concentration of surfactant (greater than
the critical micelle concentration)~\cite{coveney96}. 
These micelles tend to form in regions where the flux of particles
is low. This phenomenon is also seen when a porous medium filled with
water is flooded by a mixture of water and surfactant (fig~\ref{micell}).\\
From the upper image in fig~\ref{micell}, we can clearly see that surfactant particles
gather in the smallest pores, where the flow velocity is small. The
lower image in the
same figure represents the coloured concentration profile of
surfactant \emph{versus} time and \emph{versus} the
lattice row number (i.e. in the flow direction). This quantity is
averaged over all columns of the lattice. The start of the
simulation corresponds
to the bottom line on this diagram, at which point there is no
surfactant present. The appearance of a new colour (light blue)
corresponds to the advance of the front of
surfactant (represented approximately by the diagonal ``front'' in
this image). We can see
that the progression of the surfactant front does not exactly follow a
constant speed. Indeed, zones of high surfactant concentration appear
(represented by red peaks); these
correspond to the formation of micelles. We can see that the micelles
are stable in time, both from the point of view of their concentration
(which actually even shows a tendency to increase) and of their
localisation (the surfactant density peaks are vertical, indicating
that the micelles do not move).
The advance of surfactant is thus held up by the micellisation
process; the water advances faster than the
surfactant. 
\paragraph{Structure}
Here the structural properties of the front during imbibition
simulations are investigated.
During an invasion simulation, the precise location of the interface can be extracted. Only sites at the water-oil or surfactant-oil interface are taken into account (the obstacle
sites are discarded). From these data, the fractal dimension of the interface can be calculated.
This is achieved by counting the number of boxes needed to cover the
interface for different box sizes. The fractal dimension is the slope
of the linear part, in a log-log diagram, of the evolution of the number
of boxes \emph{versus} the size of the box (the so-called box counting method).
At the beginning of the simulation, the interface is flat, and so the
fractal dimension is equal to $1$. As invasion takes place, the
interface becomes more complex and the resulting fractal dimension
increases to reach a maximum value at the water percolation threshold.
Simulations have been performed in order to compare the fractal dimension of
the front in the binary and ternary cases. Figure~\ref{withoutpore} - bottom
shows the results obtained from several of these simulations. It can be seen
that the difference between the two curves (binary and ternary case)
is very small, implying that the fractal dimension of the front is the
same in both cases. The fractal dimension of the porous medium (equal
to $1.65$), also shown in
figure~\ref{withoutpore} - bottom, is greater than that associated with the
invading fronts.  One might thus wonder if the fractal dimension of the
front is not substantially controlled by the complexity of the porous
medium. Indeed, the linear part of these curves appears at scales
larger than $10$ lattice sites, which is greater than
the average pore size (here equal to $8.4$). The shape of the front
during invasion could be imposed by the porous medium.
Simulations of invasion in the absence of porous medium help to
resolve this issue.
Simulations with different surfactant
concentrations have been run, and the results are displayed
figure~\ref{withoutpore} - top.\\
In the case of low concentration of surfactant
(figure~\ref{withoutpore} - top) which leads to similar
results as in the absence of surfactant, the interface exhibit two
smooth bumps. On the other hand, when the concentration of surfactant
is greater ($0.3$), the interface becomes much more complicated. The calculation
of the fractal dimension in these two cases leads to the values $1.1$ and
$1.35$ respectively (lower part of figure~\ref{withoutpore}).
Simulations with higher surfactant concentrations produce the same
fractal dimensions as the latter case. Thus the fractal dimension of the invasion front
evolves from $1.0$ or $1.1$ to $1.3$ when the proportion of
surfactant changes from $0$ to $50\%$. We can see that the linear regime
in these log-log plots persists down to box sizes of a few lattice
sites, meaning that interfacial
complexity still exists at this scale. For invasion simulations in porous
media, the high values of the fractal dimensions are found at large scales, and a
decrease of the slopes is observed at smaller scales, where the
interfacial structure is not imposed by the porous medium but
rather follows the shape described in the absence of
porous medium.\\
Another, more intuitive, means of characterising the interface is simply
to compute its length (i.e. the number of
sites that it occupies). It is calculated in the same way as above by
counting the number of interfacial sites (for example a blue site is
at the interface if one of its neighbours is not a blue site nor an
obstacle site). Results are displayed in figure~\ref{lenghtinter}.
In both binary and ternary invasion, at the beginning of the
simulation, the interfacial length grows. This growth is
approximately the same in the two cases, and is roughly linear with
time over the first 20000
timesteps. No obvious reasons are apparent for this linear time
evolution. The point at which
linear growth halts corresponds approximately to the onset of
water percolation. In the binary
case, after the linear regime, the length of the interface begins to
decrease. This is related
to the fact that oil droplets do not break into smaller droplets but rather
escape from the lattice, leading to a reduction of the interfacial
length. On the other hand, in the ternary case, beyond
the linear regime, the surfactant particles induce the breaking of
large oil droplets into
smaller ones, thus increasing the overall  length of the
interface, even though the total number
of oil particles decreases as invasion drives them from the
lattice. This is due to emulsification, and to the fact that the
structure of the interface in the ternary case is much more complex.

\begin{figure}[ht]
\begin{center}
\scalebox{1.0}{\includegraphics{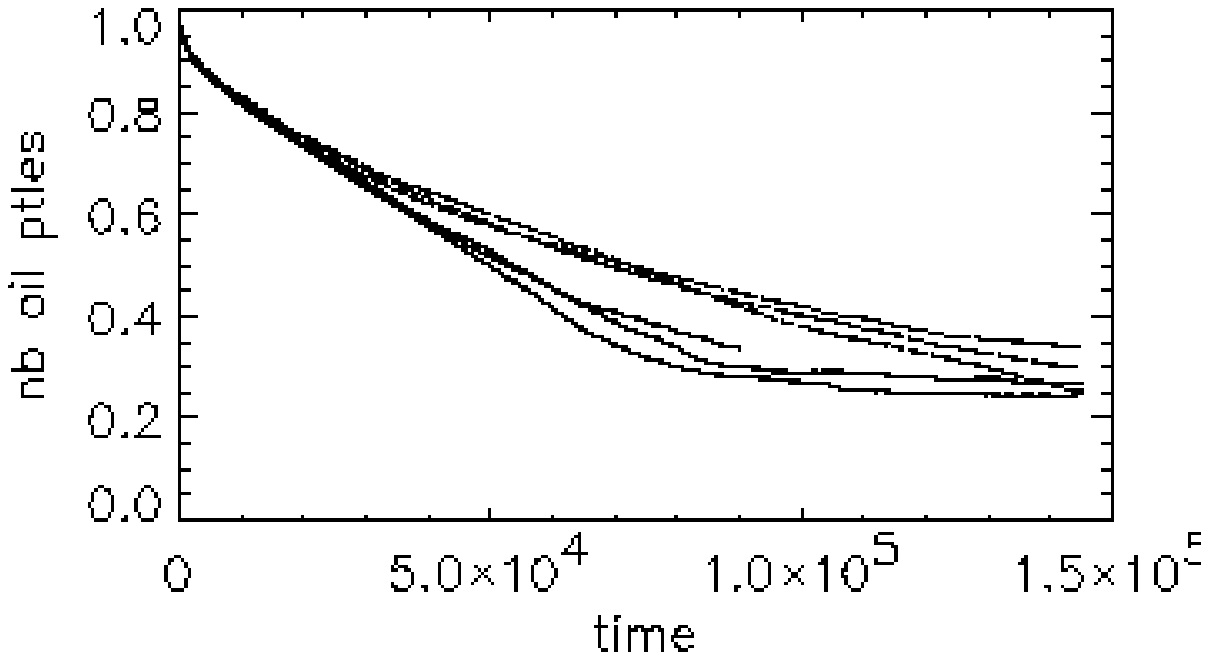}\includegraphics{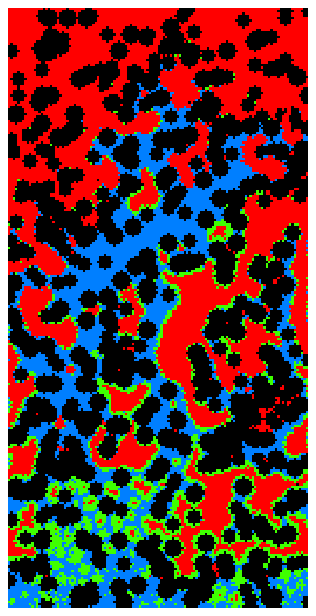}}
\end{center}
\caption{\textbf{Left: time evolution of the number of oil particles
when invading an oil-filled
water-wetting porous medium with water (three continuous lines) or with a mixture of
water and surfactant ($\rho_{surf}=0.2$)(three upper dotted lines).
The three curves in each case correspond to three independent similar
simulations. The forcing level is $0.001$. Right: a
snapshot of the case with surfactant after $70000$ timesteps.}}
\label{gellowforce}
\end{figure}

\begin{figure}[ht]
\begin{center}
\scalebox{0.7}{\includegraphics{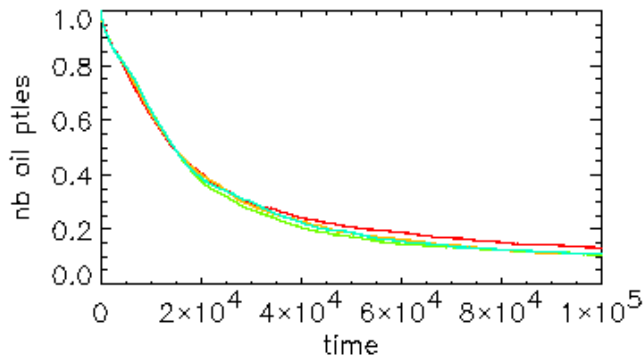}}
\end{center}
\caption{\textbf{Temporal evolution of the number of oil particles
when invading an oil-filled
porous medium (the same as previously used) with water (lower curves) or with a mixture of
water and surfactant ($\rho_{surf}=0.2$)(upper curves). The forcing
level is $0.005$ and the lattice size is $128 \times 256$. Two similar
independant simulations heve been performed in each case.}}
\label{gelhighforce}
\end{figure}

\paragraph{Timescale of micelle and emuslification phenomena}
In this part, we discuss the time scale of the phenomena described
above, due to the introduction of surfactant.\\
First of all, the formation of a complex interface appears
quickly after the beginning of the simulation, well before the onset
of water percolation.
The phenomena of micellisation and emulsification are coupled
together and appear on the same time scale.
Figure~\ref{gellowforce} shows the results obtained when
invading a porous medium filled with oil with a mixture of
surfactant and water or simply by water. Three simulations
have been run in each case. The forcing level is very small (equal to $0.001$).\\
We can see that the curves associated with invasion without surfactant
lie below the curves associated with the surfactant case. This means
that the invasion of pure water is faster. The main differences in the
curves appear near water percolation ($\sim 80000$ timesteps). After this point, the
curves corresponding to the surfactant containing fluid continue to
decrease while those associated with pure water invasion become flat.
This suggests that in the
case of pure water invasion, the oil percolation state is ended while in the case
with surfactant, particles of oil continue to flow within the medium. We
expect that the asymptotic value, that is residual oil saturation, will be
smaller in the surfactant case, due to the flow of small oil
droplets.\\
The slowing-down of the invasion process when surfactant
is present can be understood qualitatively in terms of ``gel''
formation. In the snapshot image in figure~\ref{gellowforce}, we can see a large number of
micelles and/or small oil droplets surrounded by surfactant
accumulating at the
bottom of the lattice from where invasion takes place. This
accumulation of micelles prevents the
invading fluid from passing through, slowing down the invasion
process. This is analogous to the known formation of gel by
accumulation of surfactant or polymer in actual flooding experiments
or reservoir treatments. Over long times, the effect of surfactant is to
enhance oil recovery, as can we expect from figure~\ref{gellowforce}.\\
Some additional simulations at higher
forcing levels have been run, with the same initial conditions:
the results are shown figure~\ref{gelhighforce}.
Two simulations have been run in each case (with and without
surfactant). The results with and
without surfactant present in the invading phase fluid are identical,
implying that a gel-like phase does not exist anymore;
its structure is destroyed by the greater flow rates used here.\\
In conclusion, the formation of a gel-like phase is observed at low
forcing levels, when the
invasion process is slow. Its formation is due to the self assembly of
micelles in pores.
The effect of surfactant when the forcing level is high
is small and, in some cases negligible.
From these simulations, we conclude that surfactant fluids need a
significant period of
time to produce new features through self-assembly processes. Their action
involves the formation of micelles and/or
emulsification which can then enhance oil production during the imbibition
process. However, to achieve this even at low forcing levels requires
the use of high concentrations of surfactant ($\sim 30\%$), i.e.
concentrations which would lead to microemulsion states under
equilibrium conditions.

\section{Drainage simulations}
The term ``drainage'' is used to describe the invasion of a porous
medium
filled with wetting fluid by a non-wetting one. Simulations are
performed in an oil-wetting porous medium  displayed in
figure~\ref{drainsurf}. The lattice size is $128\times 256$. The effect of the applied
force and the presence of surfactant in the invading phase are described.
\subsection{Effect of applied force}
\begin{figure}[ht]
\begin{center}
\scalebox{0.8}{\includegraphics{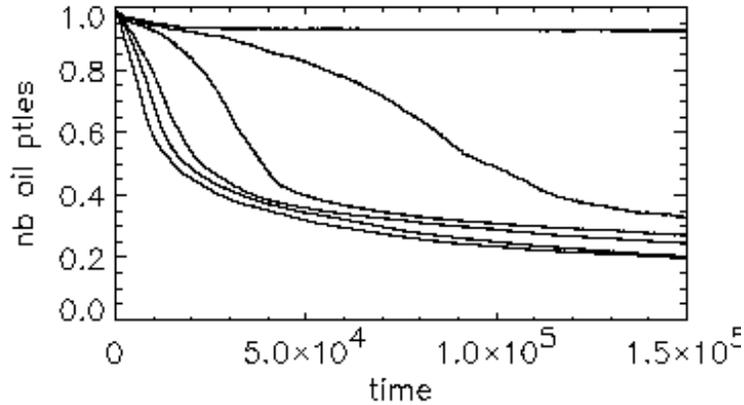}}
\end{center}
\caption{\textbf{Temporal evolution of the number of oil particles
remaining on the $128 \times 256$ lattice
when invading an oil-filled
oil-wetting porous medium with water. The various curves are at forcing
levels $0.005$, $0.01$, $0.02$, $0.03$, $0.04$, and $0.05$, from
top to bottom.}}
\label{drainforce}
\end{figure}

The methodology is the same as in our imbibition simulations
(section~\ref{imbibition}) apart from
the fact that the medium is oil-wetting.  Simulations are performed
with a wettability index of $+7$, for different fluid forcing levels. Results are
displayed in figure~\ref{drainforce}, for the binary immiscible case.\\
At very low forcing levels (upper curve), the non-wetting fluid cannot
enter the medium because of the capillary forces: the applied force on
the non-wetting fluid has to exceed the capillary force to
allow invasion. For greater driving forces, the speed of the invasion
process is roughly proportional to the applied force. As in the case
of imbibition, a maximum flow speed is reached at high forcing levels. The
residual oil saturation decreases when the applied
force increased, as previously observed in our imbibition simulations.\\
Three regimes can be distinguished during the invasion process: first,
the non-wetting fluid invades the
medium, displacing the wetting fluid. Secondly, the non-wetting fluid
percolates, but the still flowing wetting fluid retains its connectivity to the top
of the lattice. The last regime corresponds to
flow of the non-wetting fluid through a medium containing stagnant residual
wetting fluid.

\subsection{Effect of surfactant}
\begin{figure}[ht]
\begin{center}
\scalebox{1.0}{\includegraphics{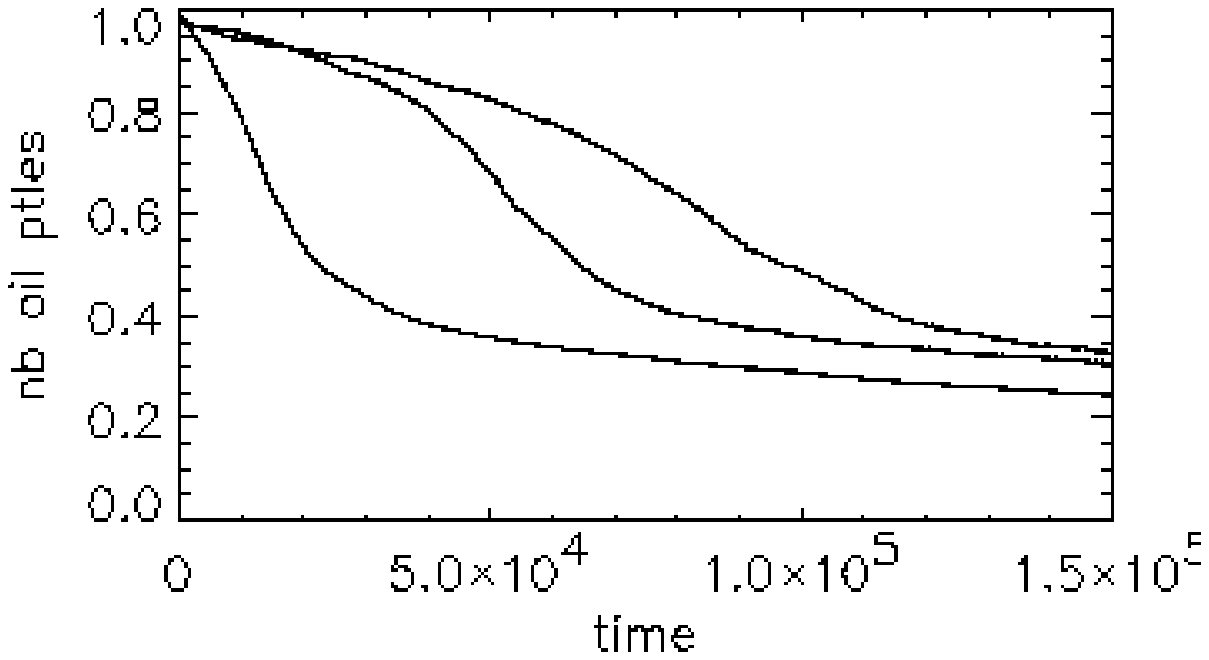}\includegraphics{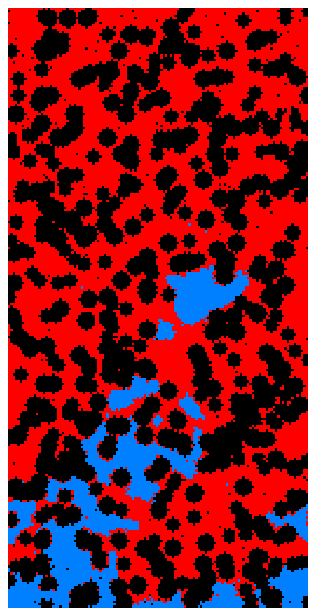}}
\end{center}
\caption{\textbf{Left: temporal evolution of the number of oil
particles remaining on a $128 \times 256$ lattice when invading an oil filled
oil-wetting porous medium with water (lower curve) or with a mixture of
water and surfactant ($15\%$ surfactant in the invading fluid
(intermediate curve) and $30\%$ (upper curve)). The forcing
level is $0.02$. Right: a snapshot of the simulation during the binary fluid
invasion process (oil in red and water in blue).}}
\label{drainsurf}
\end{figure}

\begin{figure}[p]
\begin{center}
\scalebox{0.8}{\includegraphics{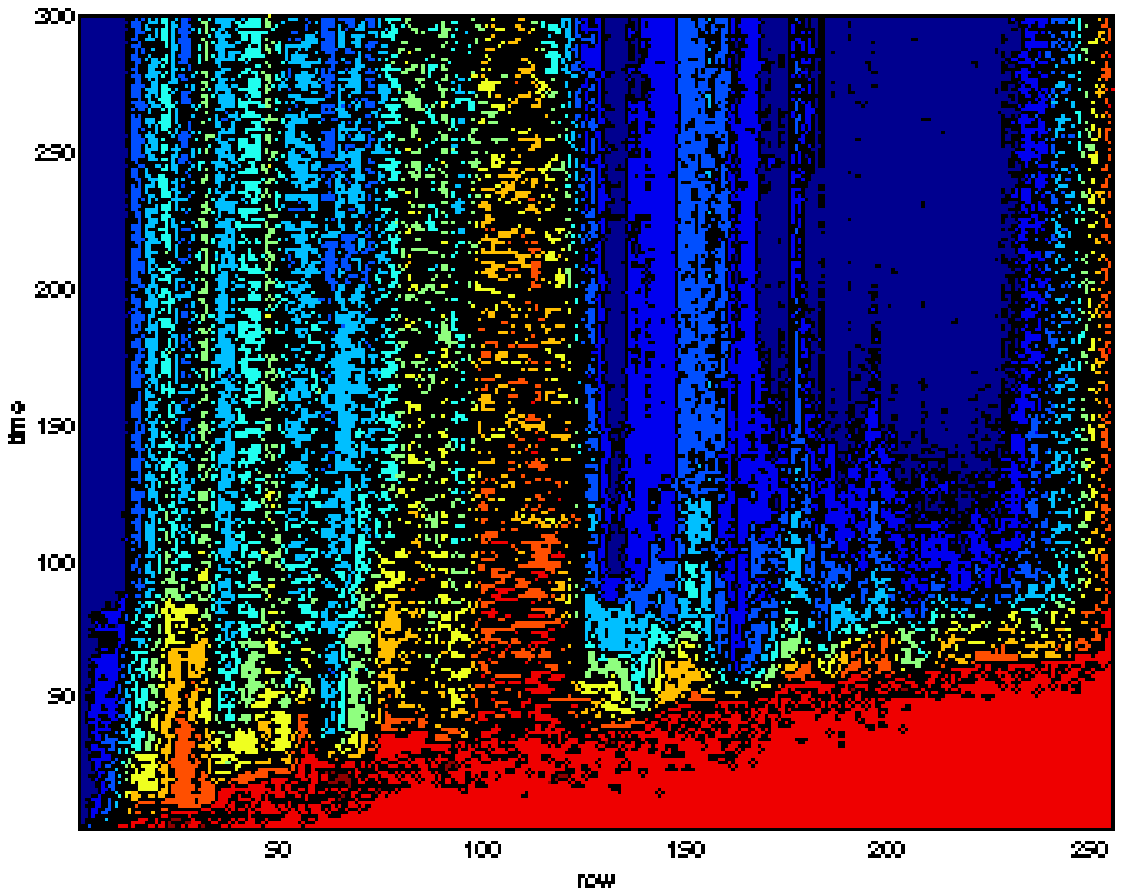}}
\scalebox{0.8}{\includegraphics{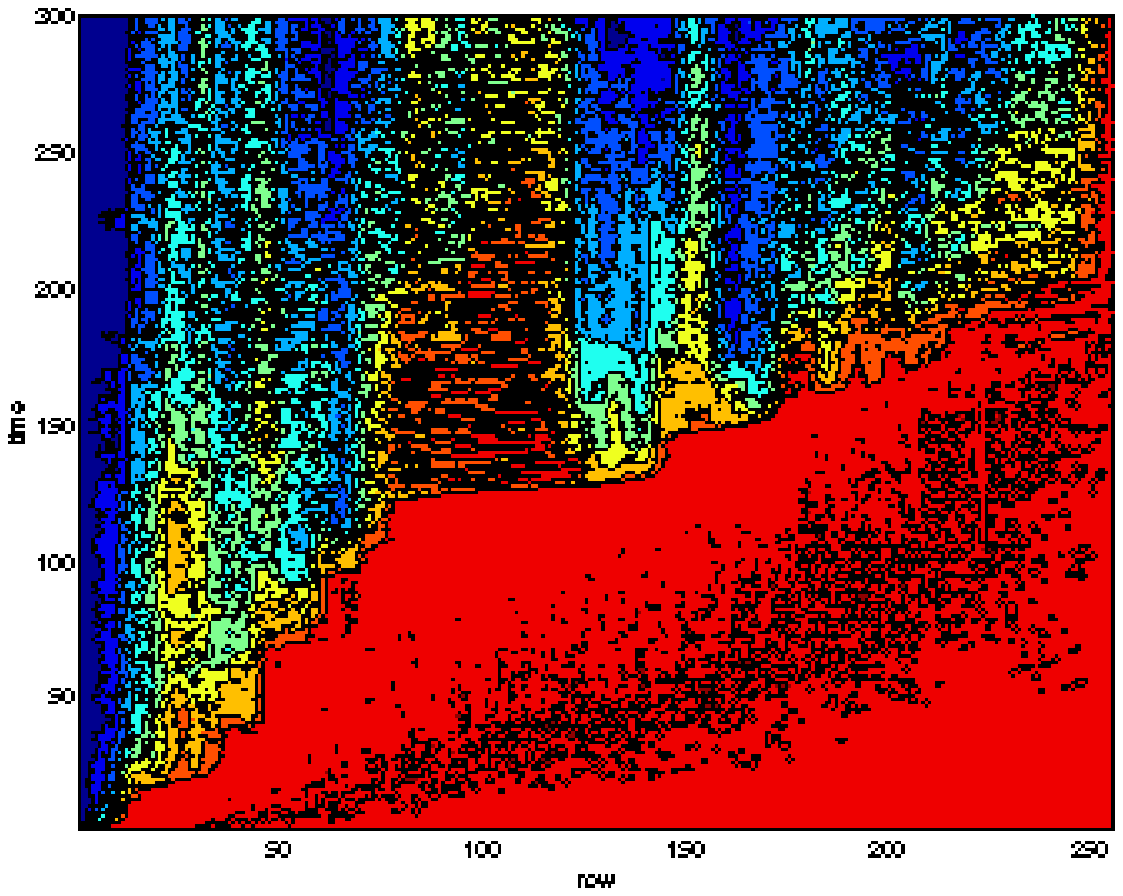}}
\end{center}
\caption{\textbf{Normalised coloured oil concentration profile (red
and blue correspond to high and low concentration respectively) \emph{versus} time
and y-coordinate (direction of the flow) in the case of drainage without
(top) and with surfactant (bottom)(at $30\%$ concentration) present in
the invading phase. The porous medium is shown in
figure~\ref{drainsurf}. One time unit corresponds to 250 timesteps.}}
\label{surconcdrain}
\end{figure}

The effect of the presence of surfactant in the invading fluid is now
investigated. Simulations have been performed with a $+7$ wettability
index (i.e. strongly oil-wetting),
a driving force of $0.02$ using the gravity condition, a reduced density of
$0.5$ and for concentrations of surfactant equal to $0$, $15\%$, and
$30\%$. Results are displayed in figure~\ref{drainsurf}.\\
We can see that the temporal evolution of the number of oil particles changes
dramatically when surfactant is added to the invading fluid. Going from
$0$ to $30\%$ surfactant concentration, the speed (or the efficiency)
of the displacement process is slowed down by more than a factor of $2$.
The residual oil saturation appears to
be lower ($\sim 5-10\%$) in the case of pure water invasion.\\
Figure~\ref{surconcdrain} displays the oil concentration profile for
drainage simulations with and without surfactant.
The red region is greater in the case with surfactant, showing that
the process is slower. Moreover, we can see that the progression of
water follows a stepwise increase. This feature is more marked in the
case with surfactant (lower part of figure~\ref{surconcdrain}). It is due to
the fact that the non-wetting
fluid experiences a delay before it is able to enter a channel.
\emph{Visual inspection reveals that the invading fluid does not take the
same path when surfactant is present as when it is absent}. In the
surfactant case, the path is much
more tortuous. In experiments, there is some indirect evidence that
paths can be different, depending on the nature of the invading fluid~\cite{olthoff}.
Our numerical results appear to confirm this behaviour.\\

\section{Conclusion}
A two-dimensional hydrodynamic lattice gas model has been used to study
the behaviour of complex multiphase and amphiphilic fluids
under various conditions in two dimensions. In simple cases, the results from this model
agree well with theoretical predictions. In more complex geometries
like porous media (where theoretical predictions in general cannot be made), an
extension of Darcy's law has been used, which explicitly admits a
viscous coupling between the fluids. This coupling appears to be
non-negligible, and exhibits a maximum for a 1:1
water and oil mixture. Such strong viscous coupling may be understood in terms of
the spatial dimensionality of the model. The introduction of
surfactant does not change the response of oil and water dramatically
but it lowers the capillary
threshold. On the other hand, during imbibition,
introduction of surfactant leads to the appearance of new and complex
features, including emulsification and micellisation. At very low
fluid forcing levels,
this leads to the production of a low-resistance gel, which then slows down the progress
of the invading fluid. At long times (beyond the water percolation threshold),
the concentration of remaining oil within the porous medium is lowered by the
action of surfactant, thus enhancing oil recovery. 
The converse behaviour is  observed in drainage simulations:
the introduction of surfactant leads to a reduction in the invasion
process and an increase in the residual oil saturation. Similar
studies are now underway using a three dimensional version of our
amphiphilic lattice gas~\cite{maillet}.

\section{Acknowledgments}
Fruitful discussions with Bruce Boghosian, James Wilson, Phillip Fowler and Omar
Al-Mushadani are gratefully acknowledged. We are grateful to Keir Novik
for his assistance in converting colour images to the greyscale
versions included here. 

\section{Appendix: Temporal \emph{versus} ensemble averaging}
In most of the simulations presented in this paper, averaging of the
fluctuations inherent in our lattice gas model has been
performed over time, when a steady state has been reached, rather than
over an ensemble of different simulations. 
When using fully periodic boundary conditions, all simulations were
started from an initially random configuration, independent initial
configurations being constructed using different seeds for the
random number generators. Assuming the existence of a steady
state, and the ergodicity of the system, we can use the ergodic
theorem to argue that
averaging in time is equivalent to ensemble averaging. 
Tests have been made concerning the equivalence
of these two procedures, and they show that the difference between
temporal and ensemble averaging is small (roughly $\sim 2.0\%$ for
medium forcing levels, although this
increases when the forcing becomes very small).\\
The results obtained in this paper can thus be compared
to those obtained using an ensemble-averaging procedure.\\
However, some thought reveals that ensemble-averaging may not always
lead to reliable results. 
For example, consider a simulation using fully periodic boundary
conditions, with a mixture of $10\%$ oil and $90\%$ water in a
water-wetting porous medium,
forcing only water, the simulation starting from an initially
random configuration. The oil particles coalesce and form a droplet
which flows. Now consider a simulation starting from a special
condition, that is a large oil bubble trapped
in the same porous medium; the probability to get such an unusual configuration from a
random initial distribution is negligible (it is essentially of zero
measure). The calculated oil flux for the bubble, whose stable
position can be found from a
preliminary imbibition simulation, will be zero. Which simulation
produces the more relevant results ? From the point of view of the
experiments, if the medium is at its residual oil saturation, the flow
of water will not induce a flow of oil and thus the appropriate
simulation is the one starting with a chosen special initial
configuration; but this has a negligible probability of being sampled
in a conventional ensemble-average. The problem with applying ensemble-averaging
to invasion simulations is that these systems are not ergodic in
general; nor indeed are their steady states equilibrium states. It is
therefore a more reliable strategy to perform and
report averages based on temporal behaviours in steady-states, as has
generally been done in this paper.\\


\newpage
\begin{thebibliography}{99}

\bibitem{frisch} U. Frisch, B. Hasslacher and Y. Pomeau (1986).
Lattice-Gas Automata for the Navier-Stokes Equation. {\sl
Physical Review Letters}, $\mathbf{56}, 1505-1508$.

\bibitem{wolfram} S. Wolfram (1986).
{\sl
J. Stat. Phys.}, $\mathbf{45}, 471-526$.

\bibitem{rothman} D. H. Rothman and J. M. Keller (1988).
Immiscible Cellular-Automaton Fluids. {\sl
J. Stat. Phys.}, $\mathbf{52}, 1119-1127$.

\bibitem{boghosian} B. M. Boghosian, P. V. Coveney and A. N. Emerton (1996).
A lattice-gas model of microemulsions.
{\sl Proc. R. Soc. Lond. A}, $\mathbf{452}, 1221-1250$.

\bibitem{boghosian99} B. M. Boghosian, P. V. Coveney and P. J. Love (1999).
Three dimensional hydrodynamic lattice gas model for amphiphilic fluid
dynamics. {\sl Proc. R. Soc. Lond. A}, in press.

\bibitem{maillet} J.-B. Maillet P. J. Love and P. V. Coveney (1999).
Three dimensional hydrodynamic lattice gas simulations of binary
immiscible and ternary amphiphilic fluid flow through porous media, in preparation.


\bibitem{wilsoncoveney} J. L. Wilson and P. V. Coveney (1997). Schlumberger internal scientific report. SCR/SR/1997/038/FCP/U.

\bibitem{kadanov} L. P. Kadanoff, G. R. McNamara and G.
Zanetti  (1987). A poiseuille viscometer for lattice gas automata. {\sl
Complex systems}, $\mathbf{1}, 791-803$.

\bibitem{coveneymaillet} P. V. Coveney, J.-B. Maillet, J. L. Wilson,
P. W. Fowler, O. Al-Mushadani and B. M. Boghosian (1998). Lattice gas
simulations of ternary amphiphilic fluid flow in porous media. {\sl
Inter. Journal of Modern Physics C}, $\mathbf{9}, 1479-1490$.

\bibitem{bookrothman} D. H. Rothman and S. Zaleski. {\em Lattice-gas
cellular automata}. (1997). Cambridge University Press.

\bibitem{rothman90} D. H. Rothman  (1990). Macroscopic laws
for immiscible two-phase flow in porous media: results from numerical simulations. {\sl
Journal of Geophysical Research}, $\mathbf{95}, 8663-8674$.

\bibitem{kalaydjian90} F. Kalaydjian (1990). Origin and quantification
of coupling between relative permeabilities for two-phase flows in
porous media. {\sl
Transport in porous media}, $\mathbf{5}, 215-229$.

\bibitem{olson97} J. F. Olson and D. H. Rothman (1997). Two-fluid flow
in sedimentary rock: simulation, transport and complexity. {\sl
J. Fluid. Mech}, $\mathbf{341}, 343-370$.

\bibitem{zarcone94} C. Zarcone and R. Lenormand (1994). D\'etermination
exp\'erimentale du couplage visqueux dans les \'ecoulements diphasiques en
milieux poreux. {\sl
C. R. Acad. Sci. Paris}, $\mathbf{318}, 1429-1435$.

\bibitem{goode93} P. A. Goode and T. S. Ramakrishnan (1993). Momentum
transfer across fluid-fluid interfaces in porous media: a network
model.
{\sl AIChE Journal}, $\mathbf{39}, 1124-1134$.

\bibitem{fower} P. Fowler and P. V. Coveney (1998). Unpublished work.

\bibitem{mailletlachet} J.-B. Maillet, V. Lachet and P. V. Coveney
(1999). Large scale molecular dynamics simulation of self-assembly
processes in short and long chain cationic surfactant.
{\sl Phys. Chem. Chem. Phys.}, in press.

\bibitem{coveney96} P. V. Coveney, A. N. Emerton and B. M. Boghosian
(1996). Simulation of self-reproducind micelles using a lattice-gas
automaton. {\sl J. Am. Chem. Soc.}, $\mathbf{118}, 10719-10724$.

\bibitem{olthoff} P. Tardy and S. Olthoff, private communication.

\end{thebibliography}
\end{document}